\documentclass[aps,pre,twocolumn,reprint,superscriptaddress]{revtex4-1}
\pdfoutput=1
\usepackage{graphicx}
\usepackage{anysize}
\usepackage{lineno}
\usepackage[amssymb]{SIunits}
\usepackage{lipsum}
\usepackage{amsmath}
\usepackage{amssymb}
\usepackage[toc,page]{appendix}
\usepackage{SIunits}
\usepackage{soul}
\usepackage{xcolor}
\usepackage{bm}
\usepackage{ulem}

\newcommand \be {\begin{equation}}
\newcommand \ee {\end{equation}}
\newcommand \bea {\begin{eqnarray}}
\newcommand \eea {\end{eqnarray}}

\newcommand \ve {\varepsilon}

\begin{document}

\title{Behavioral transition of a fish school in a crowded environment}

\author{Bruno Vent\'ejou}
\affiliation{Universit\'e Grenoble Alpes, CNRS, LIPhy, F-38000 Grenoble, France}
\author{Iris Magniez-{}-Papillon}
\affiliation{Universit\'e Grenoble Alpes, CNRS, LIPhy, F-38000 Grenoble, France}
\author{Eric Bertin}
\affiliation{Universit\'e Grenoble Alpes, CNRS, LIPhy, F-38000 Grenoble, France}
\author{Philippe Peyla}
\email[]{philippe.peyla@univ-grenoble-alpes.fr}
\affiliation{Universit\'e Grenoble Alpes, CNRS, LIPhy, F-38000 Grenoble, France}
\author{Aur\'elie Dupont}
\email[]{aurelie.dupont@univ-grenoble-alpes.fr}
\affiliation{Universit\'e Grenoble Alpes, CNRS, LIPhy, F-38000 Grenoble, France}

\begin{abstract}
In open water, social fish gather to form schools, in which fish generally align with each other. In this work, we study how this social behavior evolves when perturbed by artificial obstacles. We measure the collective behavior of a group of zebrafish in the presence of a periodic array of pillars. When pillar density is low, the fish regroup with a typical inter-distance and a well-polarized state  with parallel orientations, similar to their behavior in open water conditions. Above a critical density of pillars, their social interactions, which are mostly based on vision, are screened and the fish spread randomly through the aquarium, orienting themselves along the free axes of the pillar lattice. The abrupt transition from natural to artificial orientation happens when the pillar inter-distance is comparable to the social distance of the fish, i.e., their most probable inter-distance. We develop a stochastic model of the relative orientation between fish pairs, taking into account alignment, anti-alignment and tumbling, from a distribution biased by the environment. This model provides a good description of the experimental probability distribution of the relative orientation between the fish and captures the behavioral transition. Using the model to fit the experimental data provides qualitative information on the evolution of cognitive parameters, such as the alignment or the tumbling rates, as the pillar density increases. At high pillar density, we find that the artificial environment imposes its geometrical constraints to the fish school, drastically increasing the tumbling rate.
\end{abstract}

\maketitle

\section{Introduction}

In the animal kingdom, a large number of species live and move in groups, such as herds of mammals on land, flocks of birds in the air, and schools of fish in water \cite{Parr1927,Miller2011,Radakov1973}. 
Physical and numerical models of collective movements have shown that simple local interactions \cite{ Vicsek1995,Helbing2000} between individuals are sufficient to create the group behaviors observed in nature, like milling or schooling for fish \cite{Calovi2014}. 
Several attempts have been made to refine the models. For example, interacting neighbors can be selected by metric distance \cite{couzin_collective_2002, gregoire_moving_2003}, topological distance \cite{ballerini_interaction_2008}, Voronoi vicinity \cite{Gautrais2012, filella_model_2018}, or by taking into account the sensory information available, i.e., their field of view \cite{strandburg-peshkin_visual_2013}. 
These models accurately capture the behavior of animals in open environments without obstacles or in presence of walls in the case of fish \cite{Calovi2018} or pedestrians \cite{Helbing2000}. 

In laboratory experiments, the study of active matter in crowded environments has 
focused mainly on micro-swimmers (bacteria and micro-plankton) \cite{Nishiguchi2018,brun2019,brun2020} or synthetic 
active micro-particles \cite{Morin2017}. Indeed, micro-organisms and other self-propelled bodies in
viscous fluids are known to have complex trajectories in the presence of obstacles 
\cite{Takagi2014}. For a high density of obstacles, trapping and sub-diffusion have also been reported in a simplified theoretical model \cite{Chepizhko2013}.
These complex dynamics have garnered significant interest due to the implications for both fundamental
research and practical applications. For example, obstacles have been found to modify the 
macroscopic behavior of bacterial turbulence \cite{Nishiguchi2018}.
The presence of obstacles may also be exploited in the development of 
devices for the separation of biological cells, which plays an important role in the fields of healthcare and 
diagnostics, using techniques like the deterministic lateral displacement \cite{Hochstetter2020}.

At larger scales, e.g., for fish schools in complex environments \cite{Liao2007}, cognitive interactions, like alignment
with other fish \cite{Calovi2014}, coexist with physical interactions, such as hydrodynamic interactions at large Reynolds numbers 
\cite{Vermaa2018}. So far, experiments with fish schools have mostly been conducted in simple geometries \cite{Gautrais2012, Tunstrom2013} and the interplay between cognitive and physical interactions has mainly been addressed at a theoretical level \cite{filella_model_2018,larrieu_2021}.
However, it is worth noting that similar topics have recently been studied in the field of robotics \cite{benzion2023}. The collective behavior of a school of 
fish in a crowded environment, much like active particles, has practical applications, offering insight into the design of 
autonomous systems and robotics \cite{Ko2023}.

In the case of fish, environmental conditions can greatly influence aggregation behaviors \cite{McElroy2018}. Their natural habitat might be a small river with submerged vegetation and gravel, as for zebrafish, or reefs and rocky shores for marine species, both of which are far from the idealized open-water environment. More and more fish species are being forced to migrate due to climate change \cite{Parmesan2003}, pushing them towards environments to which they are not necessarily adapted. Individuals must use acquired environmental information to change their usual social interactions. The question of understanding how the collective behavior of a school of fish is perturbed in an obstructed environment remains open both experimentally and theoretically.

To address this question, we conducted experiments with groups of small gregarious fish, zebrafish (\textit{Danio rerio}), which are commonly used in laboratories, especially in behavioral neuroscience \cite{Miller2007}. The experiment consisted in tracking the fish positions and orientations for different densities of obstacles in the aquarium. The obstacles were opaque pillars that obstruct the fish's field of view, organized in a regular network with a $4-$fold symmetry for an optimal control over the homogeneity of pillar density. The grouping of the fish in the absence of pillars follows a log-normal distribution of the fish-to-fish distances, as previously described \cite{Becco2006}. When pillars were added, the fish inter-distance increased up to the theoretical random distribution of fish in the tank and, seemingly, to the complete screening of their social interactions.
The distribution of the fish's relative orientations was also significantly modified by the addition of pillars. A clear transition was observed from a group of mostly aligned fish to fish oriented along the axes of the pillar lattice, i.e., transition from a social order to an order imposed by the environment. 
Interestingly, the transition occured at a characteristic distance between pillars very close to the most probable fish-to-fish distance measured in the absence of obstacles. 

To further understand and characterize these observations, we propose a stochastic purely orientational model of two fish that can align, anti-align or tumble at specific rates, on top of a background angular diffusion.
The new angle resulting from a tumbling event is picked from a distribution directly obtained from the measurements and reflecting the influence of the pillars. 
The results of the model and the fitting to the experimental data are the rates of alignment, anti-alignment and tumbling for the different pillar densities, normalized by the angular diffusion coefficient.
The transition towards an orientational order imposed by the obstacles was observed again and is essentially dependent on the sharp increase in the tumbling rate. 
Hence, our model provides a qualitative evolution of ethological parameters describing the behavioral transition observed. 
In accordance with recent work by Xue et al. \cite{xue_tuning_2023}, we hypothesize
that the social interactions between fish are essentially visual, and that the progressive addition of visual obstacles severely hinder their collective behavior. Here, in contrast with previous studies \cite{lafoux_illuminance-tuned_2023,xue_tuning_2023}, the visual interaction is hindered in an anisotropic manner. Furthermore, the hydrodynamic interactions are also perturbed by the obstacles.
The observed transition illustrates the resilience of their collective behavior until a threshold where the pillars are as close as the preferred fish-to-fish distance, which brings additional discomfort.  

\section{Experimental methods}

\begin{figure}[hbt!]
    \centering
    \includegraphics[width=\linewidth]{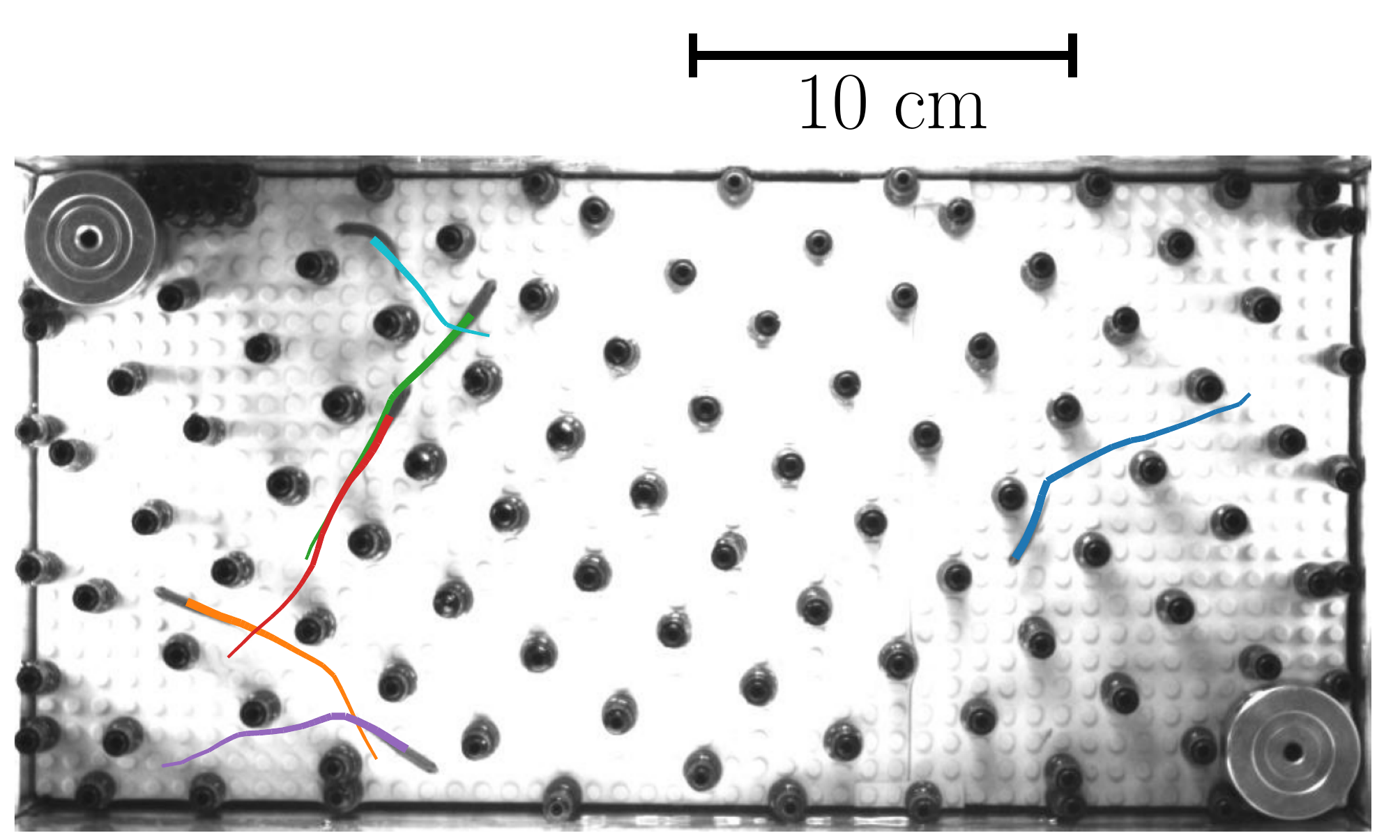}
    \caption{Snapshot of the experimental setup with recorded trajectories  of six zebrafish using TRex \cite{Walter2022}. The pillar density is $\Sigma_p = 0.12\,\textrm{cm}^{-2}$. }
    \label{fig:ExpeMethod_snapshot_d13}
\end{figure}
Zebrafish (\textit{Danio Rerio}), a freshwater fish native to South Asia, is popular as an aquarium species and also as an experimental model for research. We chose Zebrafish for their gregarious behavior associated with a tendency to be very mobile and explore the environment. Zebrafish swim by alternating burst and coast phases, allowing them to reorient their swim direction at each burst. The small size of the fish, about 3.5~cm long and 0.5~cm wide, meant that the experiments could be set up easily on the bench in a standard fish tank (36x18~cm).
From a group of 30 individuals in the resting tank, 6 were taken once a day for the experiment. The experimental tank was equipped with a Lego plate at the bottom to enable the installation of Lego pillars (diameter $0.8$~cm) at varying pillar densities (Fig.~\ref{fig:ExpeMethod_snapshot_d13}). The fish were placed in shallow water, $5$~cm deep, and in dim ambient light to limit their stress. Additional infra-red light sources were used, coupled with a Basler camera, for imaging. After a resting and adaptation period of about 20 minutes, fish behavior was recorded for 15 minutes at 20 frames per second for each trial. The fish were then put back in the resting tank. In total, about 30 different fish were used in this study.
The pillars were arranged to create a square lattice with a distance $d_p$ between the pillar centers. The minimal distance between pillars was $d_p=  1.8$~cm, leaving a space of  $1$~cm, which is slightly larger than the width of a fish, for them to be able to swim between the pillars. The maximum pillar density tested was therefore $\Sigma_p^{\textrm{max}}=0.31$ pillar.cm$^{-2}$. The fish movements were tracked using TRex \cite{Walter2022} and the obtained trajectories were further analyzed with homemade Python codes.

\begin{figure}[hbt!]
	\centering
	\includegraphics[width=\linewidth]{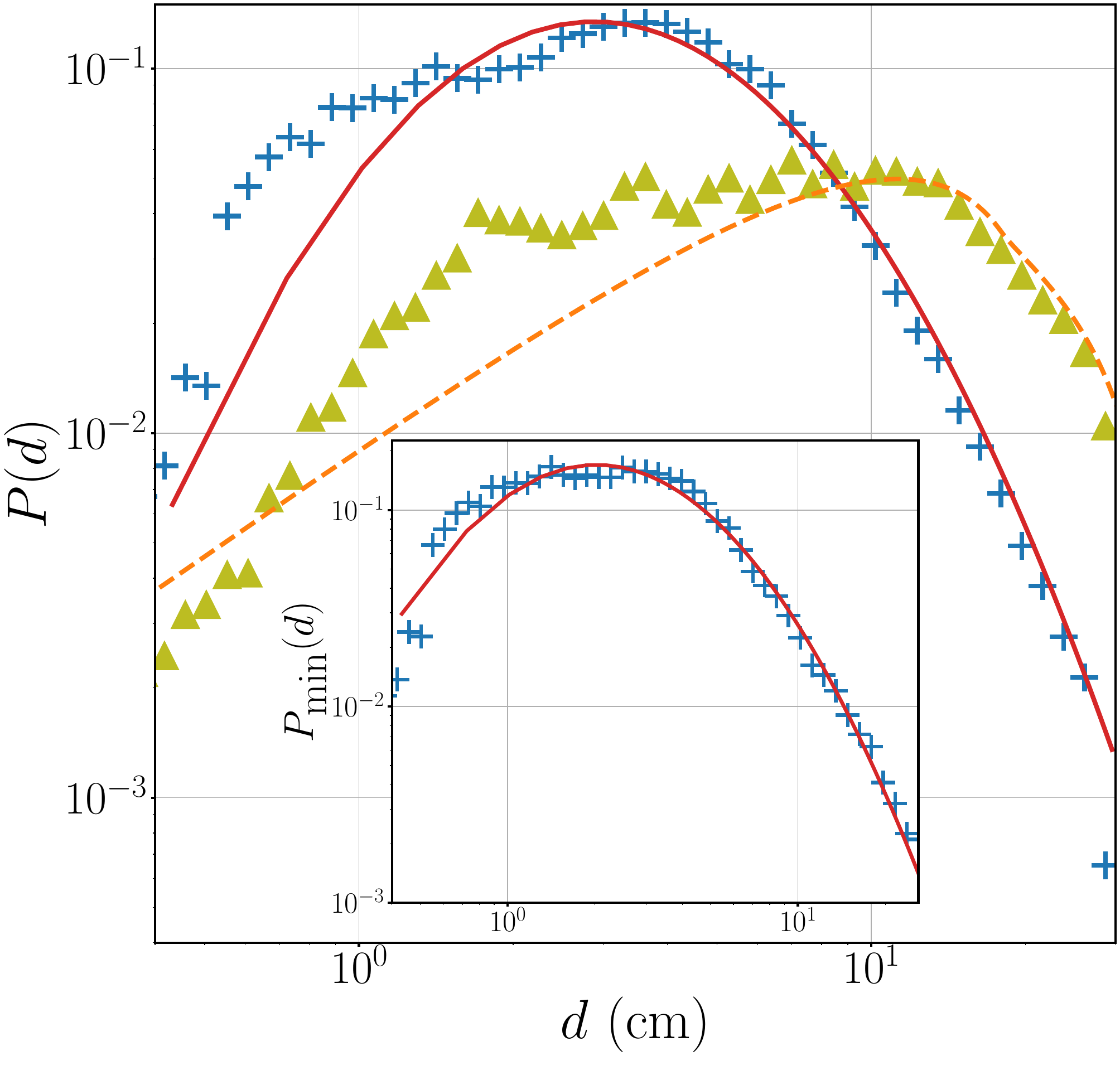}
	\caption{ Fish inter-distance probability density $P(d)$. Crosses are the experimental points without any pillars and the solid line is the log-normal fit with parameters $\mu = 1.65$ and $\sigma=0.76$. The most probable interdistance is $d_0 = 2.9\;\textrm{cm}$. The dashed line corresponds to the inter-distance probability density in the case of random points in a box. Triangles are the experimental data with the highest pillar density studied $\Sigma_p=0.31\,\textrm{cm}^{-2}$. The most probable interdistance is $d_0=7.7\pm 0.9$. Inset: Nearest-neighbor inter-distance probability density $P_{\min}(d)$ in an empty tank.  The raw data is plotted with crosses. The solid line is the log-normal fit with parameters $\mu = 1.39$ and $\sigma=0.82$.}
	\label{fig:ExpeRes_interDistance_WOPillar}
\end{figure}

\section{Experimental results}

First, the distribution $P_{\textrm{min}}(d)$ of the nearest-neighbor distance $d$ from the experimental tracks in the absence of any pillars was computed (Fig.\ref{fig:ExpeRes_interDistance_WOPillar} inset). In accordance with previous work by Becco et al. \cite{becco2006experimental}, the data are well fitted to a log-normal law as defined by:
\begin{equation}
	f(d) = \frac{1}{d \sigma \sqrt{2\pi}}\exp \left( - \frac{(\log d - \mu)^2}{2\sigma^2}\right)
\end{equation}
with fitting parameters $\mu = 1.39$ and $\sigma=0.82$.

The distribution of all fish inter-distances $d$ (not only with their nearest neighbors), denoted $P(d)$, could also be fitted to a log-normal distribution with fitting parameters $\mu = 1.65$ and $\sigma=0.76$ (Fig.~\ref{fig:ExpeRes_interDistance_WOPillar}, blue crosses and red line). The most probable inter-distance value is $d_0 =2.9\,\textrm{cm}$ and this was used later to make pillar density dimensionless. For comparison, the distribution of inter-distances in the case of randomly distributed particles in a tank of the same dimensions \cite{philip2007probability} is shown by the orange dashed line in Fig.~\ref{fig:ExpeRes_interDistance_WOPillar} and peaks at a larger distance, with the most probable value of $d_{\textrm{rand}} = 11.3 \; \textrm{cm}$.
The zebrafish observed here swam closer together in groups, highlighting the social behavior of fish schools.
Interestingly, the distribution of inter-distances in the case of the highest pillar density (green triangles in Fig.~\ref{fig:ExpeRes_interDistance_WOPillar}) almost matches the random situation. When pillar density is increased, the probability distribution changes continuously from the empty case to the random case (appendix~\ref{app:interdistance_pillarComparison}). The mean distance between fish increases with pillar density up to 
$\bar{d}=7.7 \pm 0.9\,\textrm{cm}$
which is close to the mean inter-distance in the random case. This is a first indication that the addition of obstacles to the tank tends to screen the social interactions of fish. Remarkably, the speed distribution also follows a log-normal law, as shown in the appendix \ref{app:data_cleaning}.
   
\begin{figure}[hbt!]
	\centering
	\includegraphics[width=\linewidth]{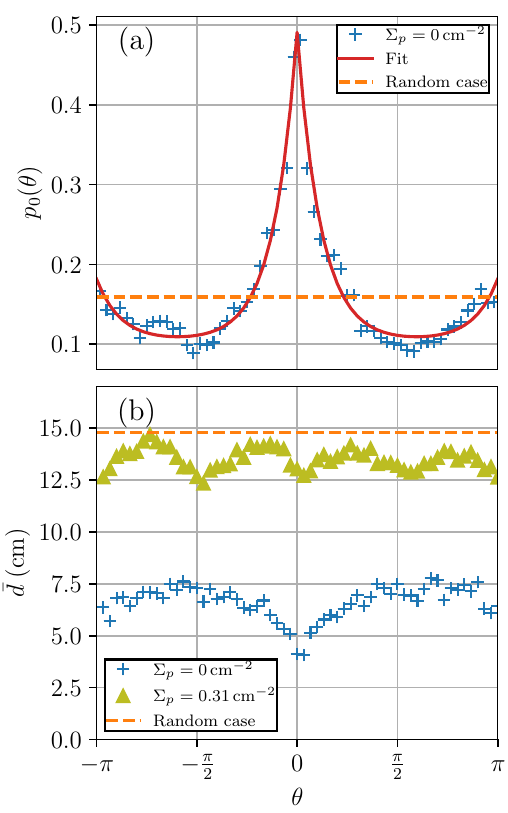}
	\caption{Fish tank without pillars. (a, crosses): $p_0(\theta)$, probability density to get angle $\theta$ between two fish. (a, dashed line): Iso-distribution attached to the random case $p_0(\theta) = \tfrac{1}{2\pi}$. (a, solid line): Exponential fit, given by Eq.~\eqref{eq_AlignAntiAlignExp}, with parameters $A_0=0.1$, $A_1=0.4$, $A_2=0.08$, $\sigma=0.4$. (b)  $\bar{d}(\theta)$, mean inter-distance (cm) between fish as a function of the relative orientation $\theta$. (b, dashed line): Mean inter-distance in the random case. (b, crosses): Mean inter-distance in the case without pillar $\Sigma_p = 0\, \textrm{cm}^{-2}$. (b, triangles): Mean inter-distance with pillar density $\Sigma_p=0.31\, \textrm{cm}^{-2}$. }
	\label{fig:ExpeRes_pdiffTheta_stateOfArt}
\end{figure}

Secondly, the distribution, $p_0(\theta)$, of the relative orientations between two fish is computed from the experimental tracks of groups of 6 fish in the absence of pillars (Figure~\ref{fig:ExpeRes_pdiffTheta_stateOfArt}(a), blue crosses). This distribution is well fitted to the function:
\begin{equation}
	p_0(\theta) = A_0 + A_1\, e^{-|\theta|/\sigma} + A_2\, e^{(\pi-|\theta|)/\sigma}\label{eq_AlignAntiAlignExp}
\end{equation}
shown as the solid red line (Fig.~\ref{fig:ExpeRes_pdiffTheta_stateOfArt}(a)). This function is made up of three terms, describing a uniform baseline, alignment and anti-alignment terms. The experimental data shows a strong alignment peak and a small anti-alignment peak. These observations are in agreement with the literature \cite{becco2006experimental} with the additional contribution of a small anti-alignment term. 

\begin{figure}[hbt!]
	\centering
	\includegraphics[width=\linewidth]{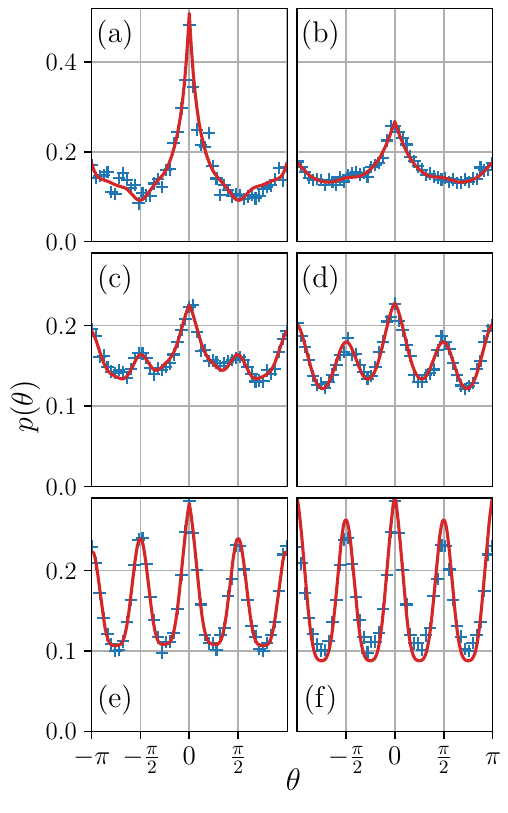}
	\caption{ Probability density distribution, $p(\theta)$, of the relative angle $\theta$ between two fish in a fish tank with pillars. Pillar density ($\text{cm}^{-2}$) are (a) $\Sigma_p=0$ (b) $\Sigma_p=0.06$. (c) $\Sigma_p=0.12$. (d) $\Sigma_p= 0.16$. (c) $\Sigma_p=0.20$. (f) $\Sigma_p=0.31$. Raw data are represented with crosses and the fits with the function given by Eq.~\eqref{eq:ExpRes_fitFunctionWithPillars} are the solid lines. }.
	\label{fig:ExpeRes_pdiffTheta}
\end{figure}

\begin{figure}[hbt!]
	\centering
	\includegraphics[width=\linewidth]{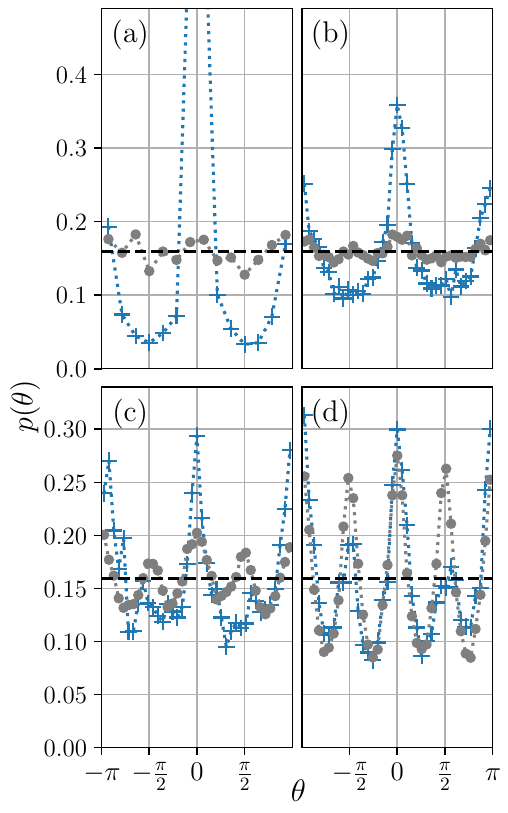}
	\caption{Fish tank with and without pillars. Probability density $p(\theta)$ to find two fish with a difference of orientation $\theta$. The crosses are corresponding to the data restricted to fish with an inter-distance $d<1.8$ cm. The dots are corresponding to the data restricted to fish with an inter-distance $d>14$ cm. (a) $\Sigma_p=0\,\textrm{cm}^{-2}$. (b) $\Sigma_p=0.06\,\textrm{cm}^{-2}$. (c) $\Sigma_p=0.16\,\textrm{cm}^{-2}$. (d) $\Sigma_p=0.31\,\textrm{cm}^{-2}$.}
	\label{fig:ExpeRes_angleDiff_InterDComparison}
\end{figure}

\begin{figure}[hbt!]
	\centering
	\includegraphics[width=\linewidth]{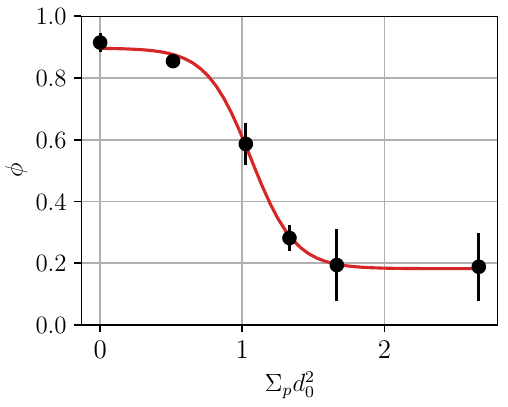}
	\caption{Order parameter, $\phi$, used to determine the transition as a function of the dimensionless pillar density $\Sigma_pd_0^2$. The inflection point given by the fit is $\Sigma_{p}^0d_0^2=1.04$. The dimensionless pillar densities from Fig.~\ref{fig:ExpeRes_pdiffTheta} are: (a) $\Sigma_pd_0^2 = 0.0$, (b) $\Sigma_pd_0^2 = 0.51$, (c) $\Sigma_pd_0^2 = 1.02$, (d) $\Sigma_pd_0^2 = 1.33$, (e) $\Sigma_pd_0^2 = 1.67$, (f) $\Sigma_pd_0^2 = 2.66$.}
	\label{fig:ExpeRes_orderParameter}
\end{figure}

We also looked at the  mean inter-distance as a function of the relative orientation between fish in the absence of pillars (Fig.~\ref{fig:ExpeRes_pdiffTheta_stateOfArt}(b), blue crosses). The distribution is symmetrical and relatively flat, except for a drop around $\theta=0$. Aligned fish are closer to each other than the overall mean inter-distance. This is confirmed by looking at the distribution $p(\theta)$ (Fig.~\ref{fig:ExpeRes_pdiffTheta}) restricted to fish pairs at two different inter-distance ranges: close to each other with an inter-distance of less than $1.8\,\textrm{cm}$ (blue crosses in Fig.~\ref{fig:ExpeRes_angleDiff_InterDComparison}(a)), or far apart with an inter-distance of more than $14\,\textrm{cm}$ (grey dots in Fig.~\ref{fig:ExpeRes_angleDiff_InterDComparison}(a)). The close fish distribution peaks sharply around $\theta=0$ and has a second smaller peak around  $\theta= \pm \pi$. Close fish are therefore mostly aligned. The distribution for fish far apart is rather flat over the whole range of angles. It is similar to an iso-distribution, as if the fish have no orientation interactions at this distance.  
These results show that the orientation interaction deteriorates with the distance between fish in the case of a small group of zebrafish, as studied herein. The relative short-range alignment interaction observed is in agreement with previous interaction models proposed for fish schools with metric distances \cite{couzin_collective_2002, gregoire_moving_2003} and with topological distances \cite{Gautrais2012, filella_model_2018}. The importance of this length scale for fish behavior has also been evidenced in the case of a fish school passing a bottleneck \cite{larrieu_fish_2023}.\\
Next, we looked at the effects of adding obstacles on the angular distribution $p(\theta)$. Figure~\ref{fig:ExpeRes_pdiffTheta} shows $p(\theta)$ for increasing pillar densities, from (a) to (f). A transition is observed from a strong peak at $\theta=0$ with a small bounce at $\theta=\pm \pi$ for a 
tank without pillars (a) to a $\tfrac{\pi}{2}$-periodic cosine distribution for the highest pillar density (f). 
The $\tfrac{\pi}{2}$ periodicity is due to  the square lattice of pillars, which allows the fish to move mainly along four directions.
	The higher the pillar density, the higher are the two intermediate peaks at $\theta=\pm \tfrac{\pi}{2}$. 
Finally, for the highest pillar density, all peaks are the same size, indicating that fish orientation is completely governed by the lattice of obstacles and no longer by social interactions.\\
As for the previous case with no pillars, the data can be fitted to a function containing symmetrical exponentials accounting for the low densities of pillars, to which periodic functions were added to account for the purely $\frac{\pi}{2}$-periodic function observed for the high densities of pillars. 
In Fig.~\ref{fig:ExpeRes_pdiffTheta}, experimental data, plotted with crosses, are fitted to the function:
\begin{align} \nonumber
	p(\theta) &= A_0 + A_1 e^{-|\theta|/\sigma} + A_2 e^{(\pi-|\theta|)/\sigma} \\
	& \qquad \qquad + B_1 \cos(4\theta) + B_2 \cos(8\theta),
	\label{eq:ExpRes_fitFunctionWithPillars}
\end{align}
which is plotted as a solid line. In order to quantify the influence of the pillar lattice on the natural alignment interaction, we define the order parameter, 
\begin{equation}
	\phi = \frac{p(0) - p(\pi/2)}{p(0)-A_0},
\end{equation}
which is related to the relative heights of the peaks at $\theta=\tfrac{\pi}{2}$ and $\theta=0$.  Hence, $\phi$ is expected to be close to one when there is no peak at $\theta=\tfrac{\pi}{2}$, and close to zero when both peaks have the same amplitude. The order parameter $\phi$, calculated from the fits of the experimental data (Fig.~\ref{fig:ExpeRes_pdiffTheta} (a) to (f)) is plotted in Fig.~\ref{fig:ExpeRes_orderParameter} as a function of the pillar density. $\phi$ decreases non-linearly from $0.9$ for the experiments with no pillars to $0.2$ for the highest pillar density. We observe a smooth transition between social alignment (high $\phi$) and obstacle-imposed orientation (low $\phi$), that can be characterized quantitatively by fitting $\phi$ with a sigmoid function and taking the pillar density at the inflection point.
The transition occurs at pillar density $\tilde{\Sigma}_p = 0.12\,\text{cm}^{-2}$ which can be converted into fish units $\tilde{\Sigma}_p d_0^2 = 1.04$. The fish unit is defined by the most probable inter-distance value between zebrafish in the absence of pillars, i.e., $d_0=2.9\,\textrm{cm}$ (see in Fig.~\ref{fig:ExpeRes_interDistance_WOPillar}). For a pillar density $\tilde{\Sigma}_p > 0.12\,\text{cm}^{-2}$, the obstacles screen the social interactions between fish. 
Again, we can split the data between fish that are close to each other, $d<1.8~\textrm{cm}$, and fish that are further apart, $d>14~\textrm{cm}$ (Fig.~\ref{fig:ExpeRes_angleDiff_InterDComparison}). For the close fish, the alignment peak is smaller than with the lowest density of pillars (b), which we interpret as being the first effects of the visual screening of fish interactions. In contrast, the long-distance order imposed by the pillar lattice appears for higher densities (c), at the order transition seen in Fig.~\ref{fig:ExpeRes_orderParameter}. Finally, both close and distant fish have the same distribution $p(\theta)$ for the situation with the most pillars in the tank (d).

To sum up, we found a correlation between the fish-to-fish distance and their relative orientation in the absence of pillars, confirming the limited range of social alignment interaction. Analysis of fish inter-distances shows a smooth transition from a group behavior at low pillar density to a quasi-random spatial distribution of fish at the highest pillar density. We observed a clear crossover between social relative orientation of fish, i.e., alignment, and environmental orientation of fish. The detailed analysis initially shows the alignment interaction screening at short distance, followed by the forced pillar order at high pillar density.  The screening of the social interactions is more obvious on the relative orientations of fish than on the fish inter-distance statistics. We will therefore focus on the relative orientation hereafter. 
\\
\section{Theoretical Model}
\label{sec:theoretical_Model}

\subsection{Definition of the model}
We consider a simple stochastic model of two swimmers that randomly align their direction of motion. The two swimmers move on a 2D plane along directions defined by the angles $\theta_1$ and $\theta_2$. During an infinitesimal time interval $[t,t+dt]$, swimmer 1 can randomly align its velocity with swimmer 2's velocity (i.e., $\theta_1 \to \theta_1'=\theta_2$) with a probability of $\lambda dt$; it can also anti-align its velocity (i.e., $\theta_1 \to \theta_1'=\theta_2+\pi$) with a probability of $\mu dt$; and it can tumble with a probability of $\nu dt$ by selecting a new angle $\theta_1'$ from a distribution $\psi(\theta_1')$, independently of $\theta_2$.
Similarly, swimmer 2 can randomly align or anti-align its velocity with that of swimmer 1, with probabilities of $\lambda dt$ and $\mu dt$ respectively; it can also tumble with probability of $\nu dt$ to a new angle $\theta_2'$ selected from the distribution $\psi(\theta_2')$.
These reorientation events are assumed to happen very quickly and are modeled as instantaneous stochastic jump processes.
Furthermore, between successive rapid reorientation events, the angles $\theta_1$ and $\theta_2$ are assumed to diffuse with a diffusion coefficient $D_R$.
The distribution $\psi(\theta')$ that models the reorientation towards the directions of the pillar lattice, is kept general at this stage. It is possible to define dimensionless ratios that can be chosen as $\tilde{\lambda} = \lambda/D_R$, $\tilde{\mu}= \mu/D_R$, and $\tilde{\nu}= \nu/D_R$.

The statistics of the angles $\theta_1$ and $\theta_2$ are described by the joint probability density function $P(\theta_1,\theta_2,t)$, which depends on time $t$. The above stochastic rules (instantaneous reorientations and angular diffusion) translate into the following evolution equation for $P(\theta_1,\theta_2,t)$:
\begin{widetext}
\bea \nonumber
	\frac{\partial P}{\partial t}(\theta_1,\theta_2,t) &=& D_R \frac{\partial^2 P}{\partial \theta_1^2}(\theta_1,\theta_2,t) + D_R \frac{\partial^2 P}{\partial \theta_2^2}(\theta_1,\theta_2,t) - 2(\lambda+\mu+\nu) P(\theta_1,\theta_2,t)\\ \nonumber
	&+& \lambda\, \delta_{2\pi}(\theta_1-\theta_2)
	\big[ P_1(\theta_1,t) + P_1(\theta_2,t)\big] 
	+ \mu\, \delta_{2\pi}(\theta_1-\theta_2-\pi)
	\big[ P_1(\theta_1,t) + P_1(\theta_2,t)\big]\\
	&+& \nu \, \psi(\theta_1) P_1(\theta_2) + \nu \, \psi(\theta_2) P_1(\theta_1),
	\label{eq:P12}
\eea
\end{widetext}  
where
\be
	P_1(\theta_1,t)=\int_{-\pi}^{\pi} d\theta_2 \, P(\theta_1,\theta_2,t)
\ee
is the marginal distribution of $\theta_1$ (or equivalently of $\theta_2$) and $\delta_{2\pi}(x) = \sum_{n=-\infty}^{\infty}\delta(x+2n\pi)$ is a generalization of the Dirac delta distribution, taking into account the $2\pi$-periodicity of angles.

\subsection{Stationary single-angle distribution $P_1 (\theta_1)$}
Starting from Eq.~\eqref{eq:P12} and integrating over $\theta_2$, we find that $P_1(\theta_1)$ obeys a closed equation
\bea\label{eq:eqDiffP1}
	&& P_1''(\theta_1) + \tilde{\mu} \big[ P_1(\theta_1-\pi)-P_1(\theta_1) \big] \\ \nonumber
	&& \qquad \qquad \qquad \qquad + \tilde{\nu} \big[ \psi(\theta_1)-P_1(\theta_1)\big] = 0.
\eea
To simplify the calculations, we assume in the following that the distribution $P_1(\theta_1 )$ is $\pi$-periodic, so that the term proportional to $\tilde{\mu}$ in Eq.~\eqref{eq:eqDiffP1} vanishes, resulting in
\be
	P_1''(\theta_1) + \tilde{\nu} \big[ \psi(\theta_1)-P_1(\theta_1)\big]=0.
	\label{eq:eqDiffP1_Final}
\ee
Although the assumption that $P_1(\theta_1 )$ is $\pi$-periodic is, in principle, an approximation, it is verified in the experimental data (see Sec.~\ref{sec:theoModel_DataAnalysis}).
We further assume that $\psi(\theta)$ is a $\pi$-periodic, even function, under a suitable choice of the origin of angles. 
These properties of $\psi(\theta)$ are related to the geometry of the pillar lattice, and are thus expected to hold in the experiment.
With these assumptions, the Fourier expansion of the distribution $\psi(\theta)$ takes the form:
\be
	\psi(\theta) = \frac{1}{2\pi} + \frac{1}{\pi} \sum_{n=1}^{\infty} \hat{\psi}_{2n} \cos \left(2n\theta\right).
	\label{eq:Fourier:psi}
\ee
The solution $P_1(\theta_1)$ of Eq.~\eqref{eq:eqDiffP1_Final} then reads
\be
	P_1(\theta_1) = \frac{1}{2\pi} + \frac{1}{\pi} \sum_{n=1}^{\infty} \hat{P}_{2n} \cos \left(2n\theta_1\right)
	\label{eq:Fourier:P1}
\ee
with
\be
	\hat{P}_{2n} = \hat{\psi}_{2n}\left(1+ \frac{4n^2}{\tilde{\nu}}\right)^{-1}.
	\label{eq:P2nPsi2n_Eq}
\ee
Eqs.~(\ref{eq:Fourier:psi}, \ref{eq:Fourier:P1}, \ref{eq:P2nPsi2n_Eq}) are used below to determine $\psi(\theta)$ from the experimental data of $P_1(\theta)$.

\subsection{Stationary angular difference distribution $p(\theta)$}

	We now consider the probability distribution $p(\theta)$ 
	of the angle difference $\theta=\theta_2-\theta_1$ defined from the stationary distribution $P_{\mathrm{st}}(\theta_1,\theta_2)$ as
	\be \label{eq:Pst:form}
	p(\theta) = \int_{-\pi}^{\pi} d\theta_1 \int_{-\pi}^{\pi} d\theta_2
	P_{\mathrm{st}}(\theta_1,\theta_2) \, \delta_{2\pi}(\theta_2-\theta_1-\theta).
	\ee
	Using the definition (\ref{eq:Pst:form}) of $p(\theta)$ into the evolution equation (\ref{eq:P12}), we obtain a second order differential equation,
	\bea
	\nonumber
	p''(\theta) \!\! &-& \!\! (\tilde{\lambda}+\tilde{\mu}+\tilde{\nu})\, p(\theta)
	+ \tilde{\lambda} \,\delta_{2\pi}(\theta)
	+ \tilde{\mu}\,\delta_{2\pi}(\theta-\pi)\\ 
	&& \quad +\tilde{\nu} \int_{-\pi}^{\pi} d\theta_2 \, \psi(\theta_2-\theta)\, P_1(\theta_2)\, = 0 .
	\label{eq:Pst:ODE}
	\eea
	Using the Fourier series expansion \eqref{eq:Fourier:psi}, we get for the integral term of Eq.~(\ref{eq:Pst:ODE})
	\bea
	&& \int_{-\pi}^{\pi} d\theta_2 \, \psi(\theta_2-\theta)\, P_1(\theta_2)\\ \nonumber
	&& \qquad \qquad \qquad = \frac{1}{2\pi} + \frac{1}{\pi} \sum_{n=1}^{\infty} \hat{\psi}_{2n} \hat{P}_{2n} \cos(2n\theta)\,.
	\eea
	As Eq.~(\ref{eq:Pst:ODE}) is invariant by changing $\theta$ into $-\theta$, $p(\theta)$ is an even function, i.e., $p(-\theta)=p(\theta)$.
	Using this property, we can simply solve Eq.~(\ref{eq:Pst:ODE}) on the open interval $(0,\pi)$, where it simplifies to
	\bea \label{eq:Pst:ODE2}
	&& p''(\theta) - (\tilde{\lambda}+\tilde{\mu}+\tilde{\nu})\, p(\theta) \\ \nonumber
	&& \qquad \qquad + \frac{\tilde{\nu}}{2\pi} + \frac{\tilde{\nu}^2}{\pi} \sum_{n=1}^{\infty} \frac{(\hat{\psi}_{2n})^2}{\tilde{\nu}+4n^2 } \, \cos(2n\theta) =0.
	\eea
	The delta functions appearing in Eq.~(\ref{eq:Pst:ODE}) are then taken into account through boundary conditions, as described below.
	We look for the solution of Eq.~(\ref{eq:Pst:ODE2}) under the form
	\bea \label{eq:gensol}
	p(\theta) = A_0 + A_1 \, e^{-s |\theta|} \!\! &+& \!\! A_2 \, e^{-s (\pi-|\theta|)}\\ \nonumber
		&+& \sum_{n=1}^{\infty} B_{2n} \cos(2n\theta)
	\eea
	where $A_i$ and $B_{2n}$ are constants to be determined, and
	\be \label{eq:def:s}
	s = \sqrt{\tilde{\lambda}+\tilde{\mu}+\tilde{\nu}}\,.
	\ee
	We thus find an equation that generalizes the empirical fitting function given in Eq.~\eqref{eq:ExpRes_fitFunctionWithPillars}. The exponential terms in Eq.~(\ref{eq:gensol}) are the two independent solutions of the homogeneous part of Eq.~(\ref{eq:Pst:ODE2}).
	The constants $A_0$ and $B_{2n}$ are determined by injecting the form
	(\ref{eq:gensol}) of $p(\theta)$ in Eq.~(\ref{eq:Pst:ODE2}), and equating to zero the constant term and the terms proportional to $\cos(2n\theta)$.
	We find
	\bea\label{eq:coefA0}
	A_0 &=& \frac{\tilde{\nu}}{2\pi(\tilde{\lambda}+\tilde{\mu}+\tilde{\nu})}\,, \\
	\label{eq:coefB}B_{2n} &=& \frac{\tilde{\nu}^2 \hat{\psi}_{2n}^2}{\pi (\tilde{\nu}+4n^2)(s^2+4n^2)}\,.
	\eea
	To determine $A_1$ and $A_2$, we proceed as follows.
	Integrating Eq.~(\ref{eq:Pst:ODE}) over the interval $[-\ve,\ve]$ and taking the limit $\ve\to 0$, one finds
	\be
	p'(0^+)-p'(0^-) + \tilde{\lambda} = 0.
	\ee
	Similarly, integrating Eq.~(\ref{eq:Pst:ODE}) over the interval $[\pi-\ve,\pi+\ve]$ and taking the limit $\ve\to 0$, we obtain
	\be
	D_R\, [p'(\pi^+)-p'(\pi^-)] + \tilde{\mu} = 0.
	\ee
	Using the $2\pi$-periodicity and parity properties of the distribution $P(\theta)$, we get
	\be
	p'(0^+) = -\frac{\tilde{\lambda}}{2}\,, \qquad
	p'(\pi^-) = \frac{\tilde{\mu}}{2}\,.
	\ee
	These two conditions are enough to determine $A_1$ and $A_2$ using Eq.~(\ref{eq:gensol}), leading to
	\bea \label{eq:coefA1}
	A_1 &=& \frac{\tilde{\lambda}\, e^{\pi s} + \tilde{\mu}}{4 s\, \sinh(\pi s)}\,,\\
	\label{eq:coefA2}A_2 &=& \frac{\tilde{\lambda} + \tilde{\mu}\, e^{\pi s}}{4 s\, \sinh(\pi s)}\,, 
	\eea
	with $s$ given in Eq.~(\ref{eq:def:s}).
	
\subsection{Data analysis} \label{sec:theoModel_DataAnalysis}
\begin{figure}[hbt!]
		\centering
		\includegraphics[width=\linewidth]{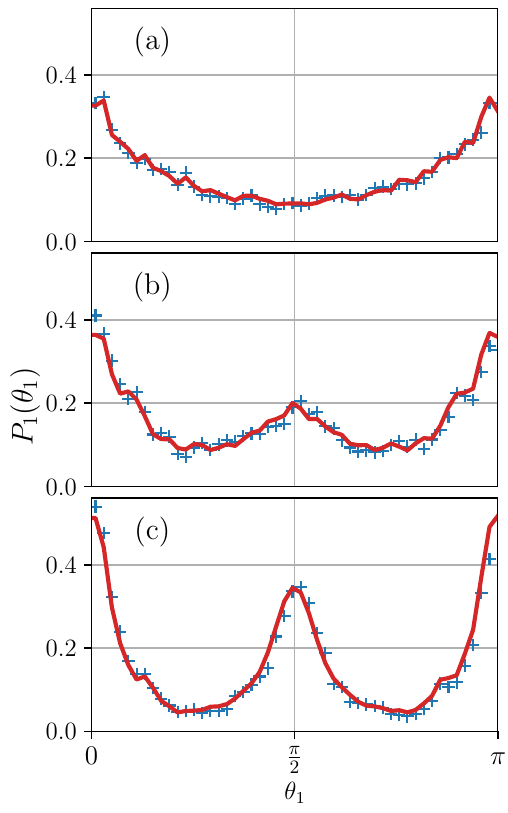}
		\caption{(a,b,c) Probability density $P_1(\theta_1)$ to find a single fish with a given orientation in a tank with pillars. The crosses represent the experimental data made symmetrical [$P_1(\theta_1)=P_1^{\mathrm{Raw}}(\theta_1) + P_1^{\mathrm{Raw}}(-\theta_1)$]. The solid line shows the expansion of $P_1(\theta)$ over the even Fourier modes. Pillar densities are (a) $\Sigma_p=0.06\textrm{cm}^{-2}$. (b) $\Sigma_p=0.16\textrm{cm}^{-2}$. (c) $\Sigma_p=0.31\,\textrm{cm}^{-2}$. }
		\label{fig:TheoModel_angleSym}
	\end{figure}

	From Eq.~\eqref{eq:gensol} and the coefficient expression Eqs.~(\ref{eq:def:s},\ref{eq:coefA0},\ref{eq:coefB},\ref{eq:coefA1},\ref{eq:coefA2}), the definition of $p(\theta)$ is fully determined by $\tilde{\lambda}$, $\tilde{\mu}$, $\tilde{\nu}$ and $B_{2n}$, where $B_{2n}$ depends only on $\tilde{\nu}$ and $\hat{P}_{2n}$ (through $\hat{\psi}_{2n}$). 
	Thus, the first step of the data analysis is to determine $\hat{P}_{2n}$ from the experimental data. 
	In Fig.~\ref{fig:TheoModel_angleSym}, the distribution $P_1(\theta)$, \textit{i.e.}, the probability of finding a fish in a given direction $\theta$, is plotted for different values of pillar density.
	The crosses show the experimental data, and the solid line is the fit of each $P_1(\theta)$ to Eq.~\eqref{eq:Fourier:P1}, using an angular Fourier expansion up to $n=11$.
	As the first orders in the angular Fourier expansion are dominant, the following data analysis is conducted with $n$ up to $n=4$.
	The good agreement between the fit and the experimental data validates the two assumptions (even function and $\pi$-periodicity) made for $P_1(\theta_1)$. %

	From Eq.~\eqref{eq:P2nPsi2n_Eq}, $\psi(\theta)$ is fully determined when $P_{2n}$ and $\tilde{\nu}$ are known. As a probability distribution, $\psi(\theta)$ must be positive by definition, and this constraint is used to reduce $\tilde{\nu}$ to a physical range designed to find $\tilde{\nu}_{\min}$ such that $\forall\,\theta\,\in\left[ -\pi;\pi\right], \; \psi^{\tilde{\nu}_{\min}}(\theta)\geq 0$. Typical probability distributions $\psi(\theta)$
	are shown in the appendix~\ref{app:psi_pillarComparison}.

	When the value $\tilde{\nu}_{\min}$ is known for  
	\begin{figure}[hbt!]
		\centering
		\includegraphics[width=\linewidth]{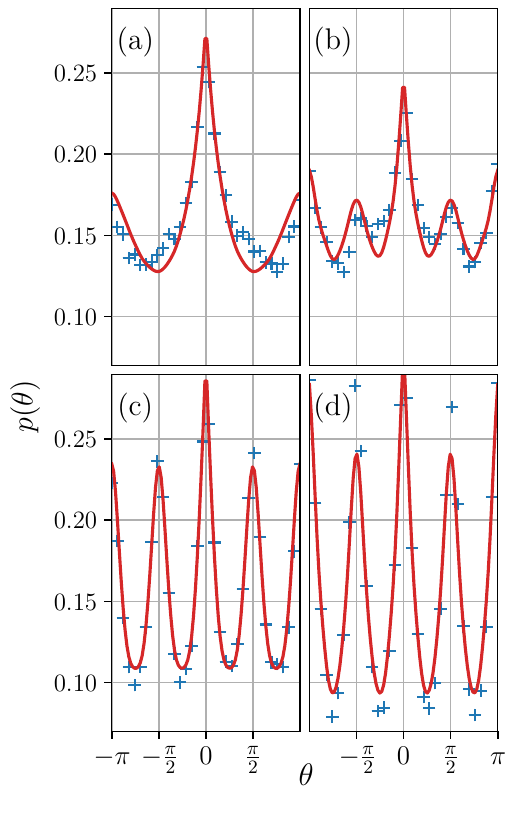}
		\caption{(a,b,c,d) Probability density $p(\theta)$ to find two fish with a relative orientation $\theta$ for increasing pillar densities from (a) to (d). The crosses are the raw data, the solid line is the fit according to the theoretical model (Eq.~\eqref{eq:gensol}). The fitting parameter values are plotted in Fig.~\ref{fig:TheoModel_lambdaMuNu}. (a) $\Sigma_p=0.06\,\textrm{cm}^{-2}$. (b) $\Sigma_p=0.12\,\textrm{cm}^{-2}$. (c) $\Sigma_p=0.20\,\textrm{cm}^{-2}$. (d) $\Sigma_p=0.31\,\textrm{cm}^{-2}$. }
		\label{fig:TheoModel_angleDiff}
	\end{figure}
	all pillar densities, Eq.~\eqref{eq:P2nPsi2n_Eq} and the coefficients $\hat{P}_{2n}$ extracted from $P_1(\theta)$ can be used to fit the experimental distribution of the relative orientations, $p(\theta)$, using Eq.~\eqref{eq:gensol}. 
	The case with no pillars is not analyzed since it is not relevant to our orientational model.
	The definition of $p(\theta)$ depends only on $\hat{P}_{2n}$ and on the three parameters $\tilde{\lambda}$, $\tilde{\mu}$, $\tilde{\nu}$, which should satisfy the constraints $\tilde{\lambda}\,\in\, \left[0;\infty\right)$, $\tilde{\mu}\,\in\, \left[0;\infty\right)$ and $\tilde{\nu}\,\in\, \left[\tilde{\nu}^{\min};\infty\right)$. 
	The results are shown in Fig.~\ref{fig:TheoModel_angleDiff}.         
	The fit is based on a dogleg algorithm, i.e., a least square method with trust region. Good agreement is generally found.
	Thus, there is a qualitative agreement between the behavior of fish and the simple stochastic description provided by alignment, anti-alignment, tumbling and  angular diffusion. \\

\begin{figure}[hbt!]
		\centering
		\includegraphics[width=\linewidth]{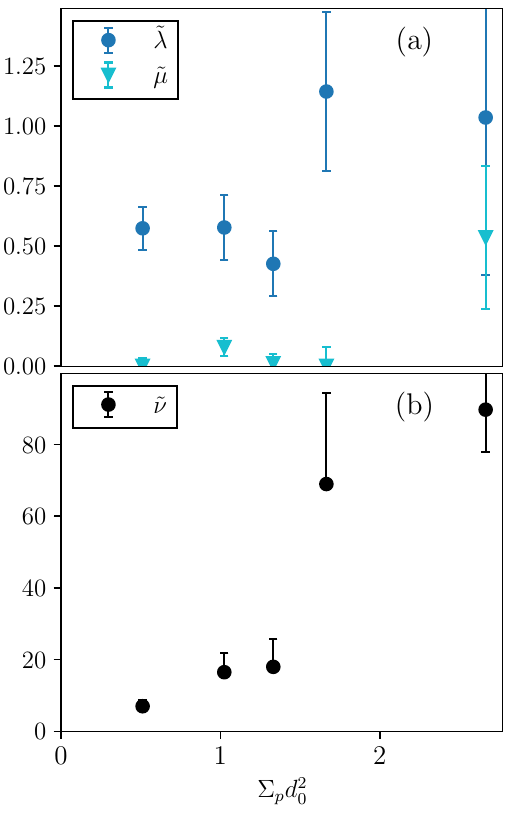}
		\caption{Fish tank with pillars. Evolution of the parameters $\tilde{\lambda}$, $\tilde{\mu}$, $\tilde{\nu}$ with the dimensionless pillar density $\Sigma_p d_0^2$. These parameters are determined by the fit done to plot the solid line in Fig.~\ref{fig:TheoModel_angleDiff} according to Eq.~\eqref{eq:gensol}. (a, Dots) Alignment parameter $\tilde{\lambda}$. (a, Triangles) Anti-alignment parameter $\tilde{\mu}$. (b) Tumbling parameter $\tilde{\nu}$.}
		\label{fig:TheoModel_lambdaMuNu}
	\end{figure}

	From the fitted data in Fig.~\ref{fig:TheoModel_angleDiff}, the evolution of $\tilde{\lambda}$ (alignment), $\tilde{\mu}$ (anti-alignment) and $\tilde{\nu}$ (tumbling) as a function of the pillar density is plotted in Fig.~\ref{fig:TheoModel_lambdaMuNu} to get a better understanding of the weight of each ingredient. 
	The error bars are large, because varying the value of the fitting parameters only has a slight effect on the goodness of the fit.
	According to Fig.~\ref{fig:TheoModel_lambdaMuNu} (a), the alignment rate is almost constant ($\tilde{\lambda} \approx 0.7 $) within the range of pillar densities tested. The anti-alignment rate, $\tilde{\mu}$, is extremely low for all pillar densities except for the largest, where $\tilde{\mu}$ attains the range of the alignment rate $\tilde{\lambda}$. The scenario is quite different for the tumbling rate, $\tilde{\nu}$, as shown in panel (b) of the same figure. After a small increase at low pillar density, the tumbling rate increases abruptly at a density of $0.20~\mathrm{cm}^{-2}$ ($\Sigma_pd^2_0 = 1.67$), almost one order of magnitude higher than the lowest value. Note that the error bars are not symmetrical due to the constraint on the minimal value $\tilde{\nu}_{min}$. The transition occurs at a larger pillar density than that characterized by the order parameter (Fig.~\ref{fig:ExpeRes_orderParameter}), for which we only quantified the ratio between observed orientational peaks. Qualitatively, we find back a smooth behavioral transition again, which is accounted for in the model by the variation of the tumbling rate. In other words, beyond a critical density of obstacles, fish reorient more often. This result is consistent with recent work by Xue et al.~\cite{xue_tuning_2023}, where a decrease in swimming burst durations and lengths was observed when the visibility was reduced by lowering the light intensity for rummy-nose tetra fish. Here, the visibility was altered by increasing the number of opaque obstacles until an apparent breakdown of the social interactions and an increased tumbling rate were reached.

\section{Conclusion}
In this work, we observed experimentally a behavioral transition of zebrafish from an aligned school to fish randomly distributed in space, and aligned with the environmental lattice. This transition from collective behavior to independent-fish behavior occurs when the density of obstacles in the aquarium is increased. The evolution is not linear with the density of obstacles, fish alignment is maintained despite the increasing difficulty to swim freely and to interact visually with congeners until a characteristic density is reached, around which a smooth change in behavior occurs. An associated characteristic length scale appears, close to the fish body length, consistent with previous work \cite{larrieu_fish_2023} that refers to this length as social distance.

The crossover is observed when the typical distance between obstacles becomes similar to the most probable fish inter-distance $d_0$ in the absence of obstacles. We propose a stochastic model to describe the observed behavioral transition. A steep increase in tumbling rate around the characteristic pillar density seems to be one of the main ingredients of the smooth transition.
 This result is consistent with recent work by Xue et al. \cite{xue_tuning_2023}, which correlated impairing the visual ability of fish with a shortening of the duration and length of their swimming bursts; it can be related to an increase in tumbling rate in our model. Our approach and other recent works \cite{lafoux_illuminance-tuned_2023,xue_tuning_2023} tackle the effects of a complex environment on the collective behavior of fish with the ultimate goal of understanding how environmental parameters, such as light, flow or obstacles, can modulate the social interactions and hence the behavior of animal groups.

\section*{Ethical statement}
Behavioral observations remained below the mild category of the severity classification of procedures, as defined by Section I of the Directive 2010/63/EU of the European Parliament and of the Council on the protection of animals used for scientific purposes. We followed the ASAB guidelines25 and ARRIVE guidelines (https://arriveguidelines.org).

\acknowledgements
This project has received financial support from ANR, through the FISHIF project, the CNRS through the MITI interdisciplinary programs. We thank our colleague P. Moreau for technical help.
\appendix
\newpage

\section{Experimental data analysis}

\subsection{Cases without pillars: Determination of bias and data cleaning}
	\subsubsection{The probability of finding a fish} \label{app:data_cleaning}
	
\begin{figure*}[hbt!]
		\centering
		\includegraphics[width=0.65\linewidth]{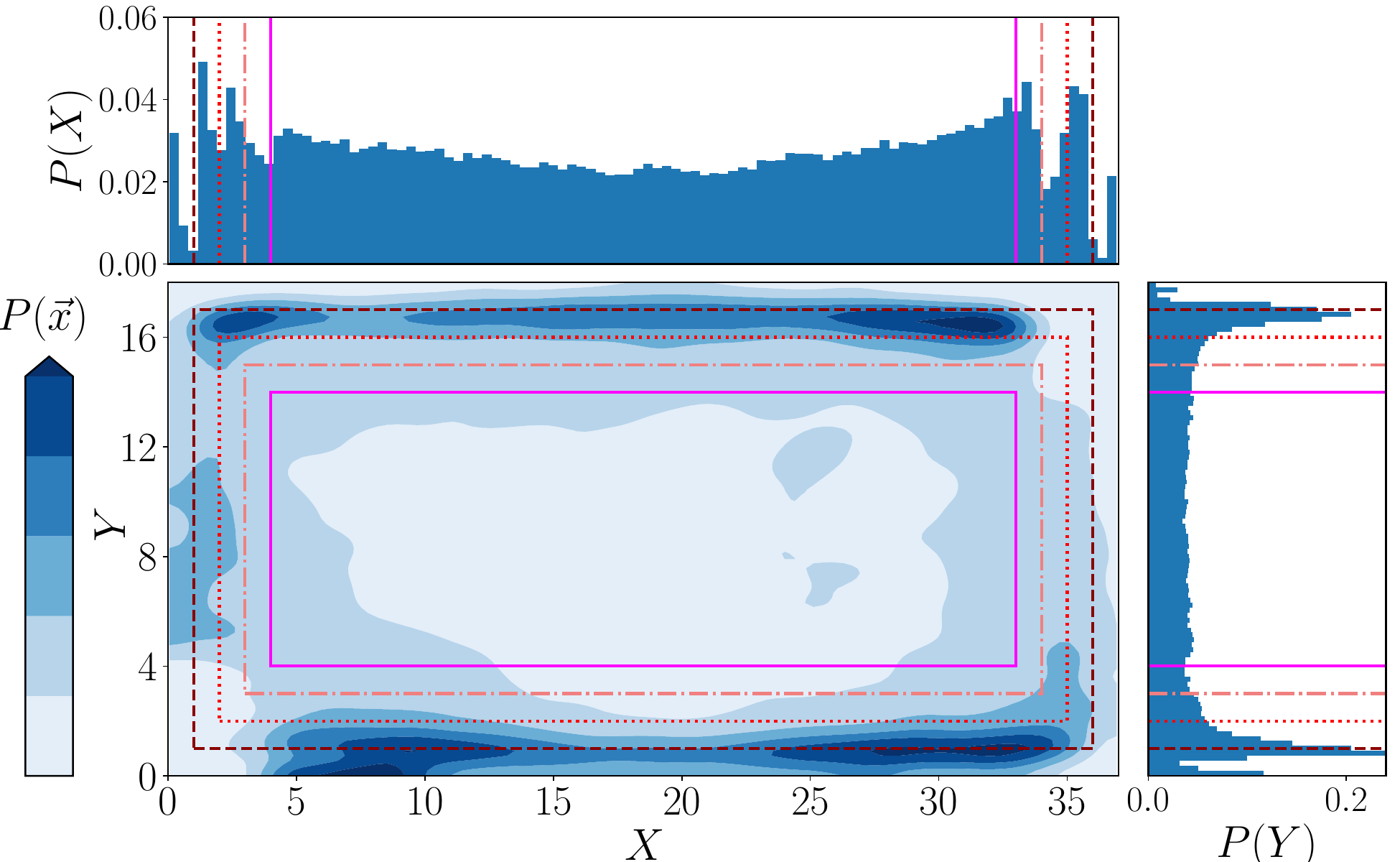}
		\caption{Fish tank without pillars. Main panel: Probability density to find a fish at a given position $\vec{x}$ averaged over 4 experiments.  Top panel: Probability density to find a fish at a given distance along the long side ($Ox$). Right panel: Probability density to find a fish at a given distance along the small side ($Oy$). The frames show the effect of cropping the data set to keep only data inside the frame.}
		\label{fig:ExpeRes_fishProbabilityDensity_emptyTank}
	\end{figure*}

	The first step of this data analysis involves determining whether any bias exists in the data set. 
	Indeed, the experiments were conducted inside a fish tank and the focus is not on the interaction between the school of fish and the tank walls but rather in the social behavior. 
The probability to find one fish at a given position in the absence of pillar is shown Fig.~\ref{fig:ExpeRes_fishProbabilityDensity_emptyTank}. The data show that fish have a tendency to follow the sides of the fish tank and it is particularly pronounced regarding the long side, highlighting the importance of the aspect ratio. 
	The probability of finding a fish on the right side is higher compare to the left side, likely influenced by the position of tank in the room. The bottom left and top right corners have a zero probability density value because large blocks anchor the Lego plate at these positions.
	To mitigate this bias, mostly visible in the absence of pillars, the data set can be restricted to specific areas  as shown by the rectangular frames in Fig.~\ref{fig:ExpeRes_fishProbabilityDensity_emptyTank}. Cropping images flattens the density probability along both sides. Still, the probability density is not perfectly uniform within the restricted area. It is important to notice that this attraction to the walls is much less prominent in the presence of pillars and hence does not influence the results.\\

	\subsubsection{Speed distribution without pillar}
\begin{figure}[hbt!]
		\centering
		\includegraphics[width=\linewidth]{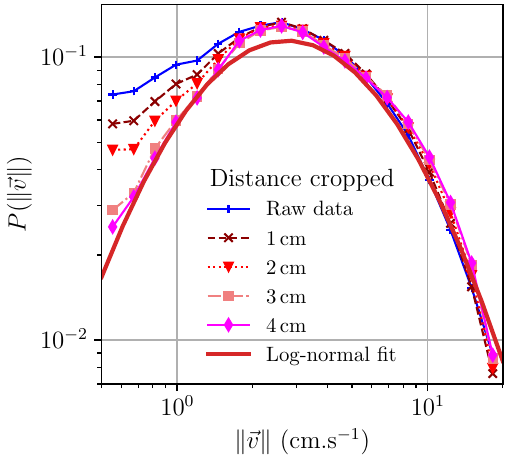}
		\caption{Fish tank without pillars. Probability density $P(||\vec{v}||)$ to find a fish at a given speed $||\vec{v}||$ averaged over 4 experiments.  (Solid line with cross): Raw data. (Other markers): Data corresponding to the crop by the same frame line as in Fig.~\ref{fig:ExpeRes_fishProbabilityDensity_emptyTank}. (Solid line): Highest crop value fitted with a log-normal law with parameters $\mu=1.766 \pm 8.10^{-3}$ and $\sigma=0.869 \pm 7.10^{-3}$.}
		\label{fig:ExpeRes_velocityProbabilityDensity_emptyTank}
	\end{figure}

	As an alternative test to determine the cropping size, the probability density of finding a fish at a given speed can be calculated and is shown in Fig.~\ref{fig:ExpeRes_velocityProbabilityDensity_emptyTank}. The larger the crop, the less likely it is to find a fish with a low speed. This implies that a significant part of immobile fish are near the tank wall. 
	Also, the density probability for the speed is close to a log-normal law for the two most cropped data set, according to the fit represented by the solid line. These two last curves are identical and it shows that a trade-off between the size of the data set and the cropped area can be found.\\

\subsection{Cases with pillars}
	\subsubsection{Interdistance with pillars} \label{app:interdistance_pillarComparison}
\begin{figure}[hbt!]
			\centering
			\includegraphics[width=7cm]{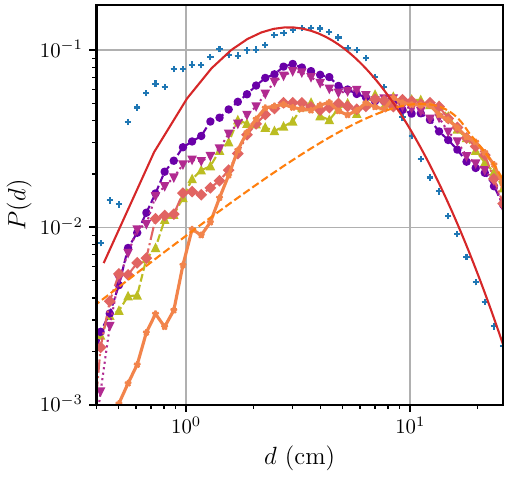}
			\caption{Fish inter-distance probability density $P(d)$ for different pillar densities values. (Crosses and solid line): Raw data without pillars $\Sigma_p = 0.0\,\textrm{cm}^{-2}$ and fits. (Dot and dashed line): $\Sigma_p = 0.06\,\textrm{cm}^{-2}$. (Down triangle and dotted line): $\Sigma_p = 0.12\,\textrm{cm}^{-2}$. (Diamond and dashed dotted line): $\Sigma_p = 0.16\,\textrm{cm}^{-2}$. (Star and solid line): $\Sigma_p = 0.20\,\textrm{cm}^{-2}$. (Up triangle and dashed line): $\Sigma_p = 0.31\,\textrm{cm}^{-2}$. (Dashed line): Interdistance in the random case.}
			\label{fig:ExpeRes_INTERDISTANCE_wPillar}
		\end{figure}

		The interdistance probability for different pillar densities is shown in Fig.~\ref{fig:ExpeRes_INTERDISTANCE_wPillar}. The solid line corresponds to the data without pillars fitted with a log-normal law, defining the reference case. The dashed line shows the case where fish are placed randomly in the tank. The interdistance probability density seems to transition continuously from the reference case to the random case when the pillar density increases. For the highest pillar density, the main difference between the data and the random case is observed for low interdistance value which could be related to bias due to interactions with the wall of the tank as mentioned above.\\
					\begin{figure}[hbt!]
			\centering
			\includegraphics[width=7cm]{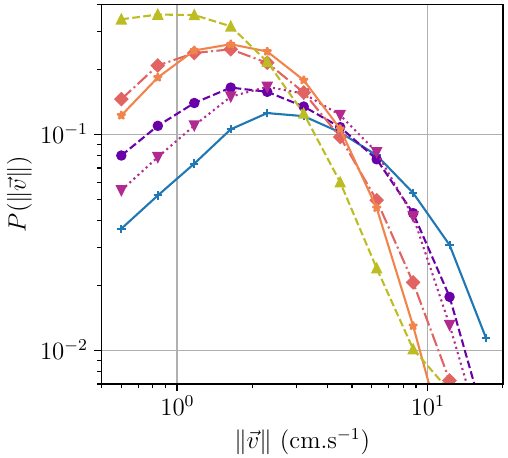}
			\caption{Probability density $P(||\vec{v}||)$ to find a fish at a given speed $||\vec{v}||$ for different pillar density values. (Crosses and solid line): $\Sigma_p = 0.0\,\textrm{cm}^{-2}$. (Dot and dashed line): $\Sigma_p = 0.06\,\textrm{cm}^{-2}$. (Down triangle and dotted line): $\Sigma_p = 0.12\,\textrm{cm}^{-2}$. (Diamond and dashed dotted line): $\Sigma_p = 0.16\,\textrm{cm}^{-2}$. (Star and solid line): $\Sigma_p = 0.20\,\textrm{cm}^{-2}$. (Up triangle and dashed line): $\Sigma_p = 0.31\,\textrm{cm}^{-2}$.}
			\label{fig:ExpeRes_velocityBias_wPillar}
		\end{figure}
		
	\subsubsection{Speed distribution with pillars}
		In Fig.~\ref{fig:ExpeRes_velocityBias_wPillar}, the probability density to find a fish at a given speed is shown for different values of pillar density. The main effect of the presence of pillars is a decrease of the average speed. It also changes the shape of the distribution, and for the highest value of pillar density, it is no more possible to use the log-normal law to describe the probability density. \\

			\begin{figure}[hbt!]
	\centering
	\includegraphics[width=6cm]{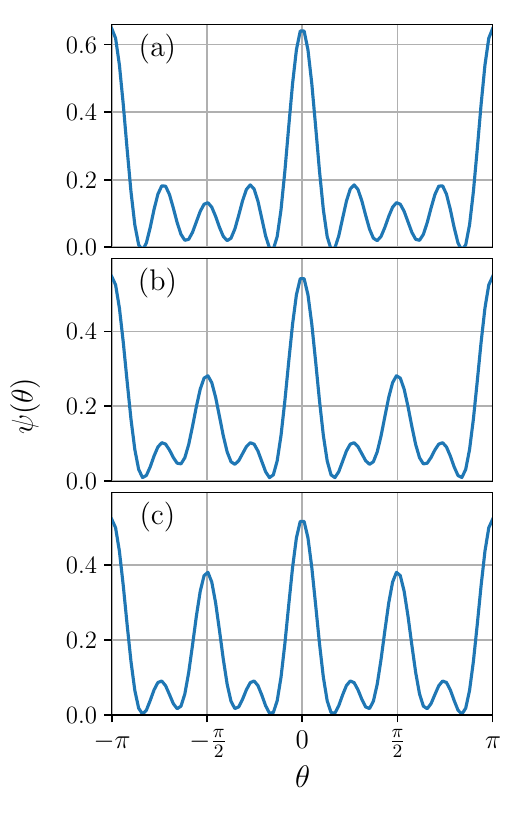}
	\caption{ Angular probability distribution $\psi(\theta)$ for the reorientation events for various pillar densities. (a) $\Sigma_p = 0.06\,\textrm{cm}^{-2}$. (b) $\Sigma_p = 0.16\,\textrm{cm}^{-2}$. (c) $\Sigma_p = 0.31\,\textrm{cm}^{-2}$. Reconstruction of the function $\psi(\theta)$ based on the value of $\tilde{\nu}$ and $\hat{P}_{2n}$ extracted from the experimental data. }
	\label{fig:TheoModel_Psi}
\end{figure}

\section{Theoretical model analysis}
\subsection*{Angular probability distribution of the reorientation event} \label{app:psi_pillarComparison}

In Fig.~\ref{fig:TheoModel_Psi}, the function $\psi(\theta)$ is shown for different values of pillar density. $\psi(\theta)$ depends only on two parameters $\tilde{\nu}$ and $\hat{P}_{2n}$, where $\hat{P}_{2n}$ are determined from the experimental data (Fig.~\ref{fig:TheoModel_angleSym}). The constraint $\psi(\theta) \geq 0 $ is translated into a constraint on $\tilde{\nu}$ such that $\tilde{\nu} \geq \tilde{\nu}^{min}$. When $\tilde{\nu}^{min}$ is determined, the function $p(\theta)$ is fitted according to Eq.~\eqref{eq:gensol} and constraints $\tilde{\lambda}\,\in\, \left[0;\infty\right)$, $\tilde{\mu}\,\in\, \left[0;\infty\right)$ and $\tilde{\nu}\,\in\, \left[\tilde{\nu}^{\min};\infty\right)$. 
Thus, the fit provides a value for $\tilde{\nu}$. The function $\psi(\theta)$ is plotted in Fig.~\ref{fig:TheoModel_Psi} using $\tilde{\nu}$ (fit) and $\hat{P}_{2n}$ (experimental data). 
%
	

\begin{thebibliography}{36}%
	\makeatletter
	\providecommand \@ifxundefined [1]{%
	 \@ifx{#1\undefined}
	}%
	\providecommand \@ifnum [1]{%
	 \ifnum #1\expandafter \@firstoftwo
	 \else \expandafter \@secondoftwo
	 \fi
	}%
	\providecommand \@ifx [1]{%
	 \ifx #1\expandafter \@firstoftwo
	 \else \expandafter \@secondoftwo
	 \fi
	}%
	\providecommand \natexlab [1]{#1}%
	\providecommand \enquote  [1]{``#1''}%
	\providecommand \bibnamefont  [1]{#1}%
	\providecommand \bibfnamefont [1]{#1}%
	\providecommand \citenamefont [1]{#1}%
	\providecommand \href@noop [0]{\@secondoftwo}%
	\providecommand \href [0]{\begingroup \@sanitize@url \@href}%
	\providecommand \@href[1]{\@@startlink{#1}\@@href}%
	\providecommand \@@href[1]{\endgroup#1\@@endlink}%
	\providecommand \@sanitize@url [0]{\catcode `\\12\catcode `\$12\catcode
	  `\&12\catcode `\#12\catcode `\^12\catcode `\_12\catcode `\%12\relax}%
	\providecommand \@@startlink[1]{}%
	\providecommand \@@endlink[0]{}%
	\providecommand \url  [0]{\begingroup\@sanitize@url \@url }%
	\providecommand \@url [1]{\endgroup\@href {#1}{\urlprefix }}%
	\providecommand \urlprefix  [0]{URL }%
	\providecommand \Eprint [0]{\href }%
	\providecommand \doibase [0]{http://dx.doi.org/}%
	\providecommand \selectlanguage [0]{\@gobble}%
	\providecommand \bibinfo  [0]{\@secondoftwo}%
	\providecommand \bibfield  [0]{\@secondoftwo}%
	\providecommand \translation [1]{[#1]}%
	\providecommand \BibitemOpen [0]{}%
	\providecommand \bibitemStop [0]{}%
	\providecommand \bibitemNoStop [0]{.\EOS\space}%
	\providecommand \EOS [0]{\spacefactor3000\relax}%
	\providecommand \BibitemShut  [1]{\csname bibitem#1\endcsname}%
	\let\auto@bib@innerbib\@empty
	\bibitem [{\citenamefont {Parr}(1927)}]{Parr1927}%
	  \BibitemOpen
	  \bibfield  {author} {\bibinfo {author} {\bibfnamefont {A.~E.}\ \bibnamefont
	  {Parr}},\ }\href@noop {} {\bibfield  {journal} {\bibinfo  {journal} {Occas.
	  Pap. Bigham Ocean- ogr. Coil.}\ }\textbf {\bibinfo {volume} {1}},\ \bibinfo
	  {pages} {1} (\bibinfo {year} {1927})}\BibitemShut {NoStop}%
	\bibitem [{\citenamefont {Miller}\ and\ \citenamefont
	  {Gerlai}(2011)}]{Miller2011}%
	  \BibitemOpen
	  \bibfield  {author} {\bibinfo {author} {\bibfnamefont {N.~Y.}\ \bibnamefont
	  {Miller}}\ and\ \bibinfo {author} {\bibfnamefont {R.}~\bibnamefont
	  {Gerlai}},\ }\href@noop {} {\bibfield  {journal} {\bibinfo  {journal} {Rev.
	  Neurosci.}\ }\textbf {\bibinfo {volume} {22}},\ \bibinfo {pages} {17}
	  (\bibinfo {year} {2011})}\BibitemShut {NoStop}%
	\bibitem [{\citenamefont {Radakov}(1973)}]{Radakov1973}%
	  \BibitemOpen
	  \bibfield  {author} {\bibinfo {author} {\bibfnamefont {D.~V.}\ \bibnamefont
	  {Radakov}},\ }\href@noop {} {\emph {\bibinfo {title} {Schooling in the
	  Ecology of Fish}}}\ (\bibinfo  {publisher} {John Wiley New York},\ \bibinfo
	  {year} {1973})\BibitemShut {NoStop}%
	\bibitem [{\citenamefont {Vicsek}\ \emph {et~al.}(1995)\citenamefont {Vicsek},
	  \citenamefont {Czirok}, \citenamefont {Ben-Jacob}, \citenamefont {Cohen},\
	  and\ \citenamefont {Shochet}}]{Vicsek1995}%
	  \BibitemOpen
	  \bibfield  {author} {\bibinfo {author} {\bibfnamefont {T.}~\bibnamefont
	  {Vicsek}}, \bibinfo {author} {\bibfnamefont {A.}~\bibnamefont {Czirok}},
	  \bibinfo {author} {\bibfnamefont {E.}~\bibnamefont {Ben-Jacob}}, \bibinfo
	  {author} {\bibfnamefont {I.}~\bibnamefont {Cohen}}, \ and\ \bibinfo {author}
	  {\bibfnamefont {O.}~\bibnamefont {Shochet}},\ }\href@noop {} {\bibfield
	  {journal} {\bibinfo  {journal} {Phys. Rev. Lett.}\ }\textbf {\bibinfo
	  {volume} {75}},\ \bibinfo {pages} {1226} (\bibinfo {year}
	  {1995})}\BibitemShut {NoStop}%
	\bibitem [{\citenamefont {Helbing}\ \emph {et~al.}(2000)\citenamefont
	  {Helbing}, \citenamefont {Farkas},\ and\ \citenamefont
	  {Vicsek}}]{Helbing2000}%
	  \BibitemOpen
	  \bibfield  {author} {\bibinfo {author} {\bibfnamefont {D.}~\bibnamefont
	  {Helbing}}, \bibinfo {author} {\bibfnamefont {I.}~\bibnamefont {Farkas}}, \
	  and\ \bibinfo {author} {\bibfnamefont {T.}~\bibnamefont {Vicsek}},\
	  }\href@noop {} {\bibfield  {journal} {\bibinfo  {journal} {Nature}\ }\textbf
	  {\bibinfo {volume} {407}},\ \bibinfo {pages} {487} (\bibinfo {year}
	  {2000})}\BibitemShut {NoStop}%
	\bibitem [{\citenamefont {Calovi}\ \emph {et~al.}(2014)\citenamefont {Calovi},
	  \citenamefont {Lopez}, \citenamefont {Ngo}, \citenamefont {Sire},
	  \citenamefont {Chat\'e},\ and\ \citenamefont {Theraulaz}}]{Calovi2014}%
	  \BibitemOpen
	  \bibfield  {author} {\bibinfo {author} {\bibfnamefont {D.~S.}\ \bibnamefont
	  {Calovi}}, \bibinfo {author} {\bibfnamefont {U.}~\bibnamefont {Lopez}},
	  \bibinfo {author} {\bibfnamefont {S.}~\bibnamefont {Ngo}}, \bibinfo {author}
	  {\bibfnamefont {C.}~\bibnamefont {Sire}}, \bibinfo {author} {\bibfnamefont
	  {H.}~\bibnamefont {Chat\'e}}, \ and\ \bibinfo {author} {\bibfnamefont
	  {G.}~\bibnamefont {Theraulaz}},\ }\href@noop {} {\bibfield  {journal}
	  {\bibinfo  {journal} {New J. Phys.}\ }\textbf {\bibinfo {volume} {16}},\
	  \bibinfo {pages} {015026} (\bibinfo {year} {2014})}\BibitemShut {NoStop}%
	\bibitem [{\citenamefont {Couzin}\ \emph {et~al.}(2002)\citenamefont {Couzin},
	  \citenamefont {Krause}, \citenamefont {James}, \citenamefont {Ruxton},\ and\
	  \citenamefont {Franks}}]{couzin_collective_2002}%
	  \BibitemOpen
	  \bibfield  {author} {\bibinfo {author} {\bibfnamefont {I.~D.}\ \bibnamefont
	  {Couzin}}, \bibinfo {author} {\bibfnamefont {J.}~\bibnamefont {Krause}},
	  \bibinfo {author} {\bibfnamefont {R.}~\bibnamefont {James}}, \bibinfo
	  {author} {\bibfnamefont {G.~D.}\ \bibnamefont {Ruxton}}, \ and\ \bibinfo
	  {author} {\bibfnamefont {N.~R.}\ \bibnamefont {Franks}},\ }\href {\doibase
	  10.1006/jtbi.2002.3065} {\bibfield  {journal} {\bibinfo  {journal} {J. Theor.
	  Biol.}\ }\textbf {\bibinfo {volume} {218}},\ \bibinfo {pages} {1} (\bibinfo
	  {year} {2002})}\BibitemShut {NoStop}%
	\bibitem [{\citenamefont {Grégoire}\ \emph {et~al.}(2003)\citenamefont
	  {Grégoire}, \citenamefont {Chaté},\ and\ \citenamefont
	  {Tu}}]{gregoire_moving_2003}%
	  \BibitemOpen
	  \bibfield  {author} {\bibinfo {author} {\bibfnamefont {G.}~\bibnamefont
	  {Grégoire}}, \bibinfo {author} {\bibfnamefont {H.}~\bibnamefont {Chaté}}, \
	  and\ \bibinfo {author} {\bibfnamefont {Y.}~\bibnamefont {Tu}},\ }\href
	  {\doibase 10.1016/S0167-2789(03)00102-7} {\bibfield  {journal} {\bibinfo
	  {journal} {Physica D}\ }\textbf {\bibinfo {volume} {181}},\ \bibinfo {pages}
	  {157} (\bibinfo {year} {2003})}\BibitemShut {NoStop}%
	\bibitem [{\citenamefont {Ballerini}\ \emph {et~al.}(2008)\citenamefont
	  {Ballerini}, \citenamefont {Cabibbo}, \citenamefont {Candelier},
	  \citenamefont {Cavagna}, \citenamefont {Cisbani}, \citenamefont {Giardina},
	  \citenamefont {Lecomte}, \citenamefont {Orlandi}, \citenamefont {Parisi},
	  \citenamefont {Procaccini}, \citenamefont {Viale},\ and\ \citenamefont
	  {Zdravkovic}}]{ballerini_interaction_2008}%
	  \BibitemOpen
	  \bibfield  {author} {\bibinfo {author} {\bibfnamefont {M.}~\bibnamefont
	  {Ballerini}}, \bibinfo {author} {\bibfnamefont {N.}~\bibnamefont {Cabibbo}},
	  \bibinfo {author} {\bibfnamefont {R.}~\bibnamefont {Candelier}}, \bibinfo
	  {author} {\bibfnamefont {A.}~\bibnamefont {Cavagna}}, \bibinfo {author}
	  {\bibfnamefont {E.}~\bibnamefont {Cisbani}}, \bibinfo {author} {\bibfnamefont
	  {I.}~\bibnamefont {Giardina}}, \bibinfo {author} {\bibfnamefont
	  {V.}~\bibnamefont {Lecomte}}, \bibinfo {author} {\bibfnamefont
	  {A.}~\bibnamefont {Orlandi}}, \bibinfo {author} {\bibfnamefont
	  {G.}~\bibnamefont {Parisi}}, \bibinfo {author} {\bibfnamefont
	  {A.}~\bibnamefont {Procaccini}}, \bibinfo {author} {\bibfnamefont
	  {M.}~\bibnamefont {Viale}}, \ and\ \bibinfo {author} {\bibfnamefont
	  {V.}~\bibnamefont {Zdravkovic}},\ }\href {\doibase 10.1073/pnas.0711437105}
	  {\bibfield  {journal} {\bibinfo  {journal} {Proc. Nat. Acad. Sci. USA}\
	  }\textbf {\bibinfo {volume} {105}},\ \bibinfo {pages} {1232} (\bibinfo {year}
	  {2008})}\BibitemShut {NoStop}%
	\bibitem [{\citenamefont {Gautrais}\ \emph {et~al.}(2012)\citenamefont
	  {Gautrais}, \citenamefont {Ginelli}, \citenamefont {Fournier}, \citenamefont
	  {Blanco}, \citenamefont {Soria}, \citenamefont {Chat\'e},\ and\ \citenamefont
	  {Theraulaz}}]{Gautrais2012}%
	  \BibitemOpen
	  \bibfield  {author} {\bibinfo {author} {\bibfnamefont {J.}~\bibnamefont
	  {Gautrais}}, \bibinfo {author} {\bibfnamefont {F.}~\bibnamefont {Ginelli}},
	  \bibinfo {author} {\bibfnamefont {R.}~\bibnamefont {Fournier}}, \bibinfo
	  {author} {\bibfnamefont {S.}~\bibnamefont {Blanco}}, \bibinfo {author}
	  {\bibfnamefont {M.}~\bibnamefont {Soria}}, \bibinfo {author} {\bibfnamefont
	  {H.}~\bibnamefont {Chat\'e}}, \ and\ \bibinfo {author} {\bibfnamefont
	  {G.}~\bibnamefont {Theraulaz}},\ }\href@noop {} {\bibfield  {journal}
	  {\bibinfo  {journal} {PLoS Comput. Biol.}\ }\textbf {\bibinfo {volume} {8}},\
	  \bibinfo {pages} {e1002678} (\bibinfo {year} {2012})}\BibitemShut {NoStop}%
	\bibitem [{\citenamefont {Filella}\ \emph {et~al.}(2018)\citenamefont
	  {Filella}, \citenamefont {Nadal}, \citenamefont {Sire}, \citenamefont
	  {Kanso},\ and\ \citenamefont {Eloy}}]{filella_model_2018}%
	  \BibitemOpen
	  \bibfield  {author} {\bibinfo {author} {\bibfnamefont {A.}~\bibnamefont
	  {Filella}}, \bibinfo {author} {\bibfnamefont {F.}~\bibnamefont {Nadal}},
	  \bibinfo {author} {\bibfnamefont {C.}~\bibnamefont {Sire}}, \bibinfo {author}
	  {\bibfnamefont {E.}~\bibnamefont {Kanso}}, \ and\ \bibinfo {author}
	  {\bibfnamefont {C.}~\bibnamefont {Eloy}},\ }\href {\doibase
	  10.1103/PhysRevLett.120.198101} {\bibfield  {journal} {\bibinfo  {journal}
	  {Phys. Rev. Lett.}\ }\textbf {\bibinfo {volume} {120}},\ \bibinfo {pages}
	  {198101} (\bibinfo {year} {2018})}\BibitemShut {NoStop}%
	\bibitem [{\citenamefont {Strandburg-Peshkin}\ \emph
	  {et~al.}(2013)\citenamefont {Strandburg-Peshkin}, \citenamefont {Twomey},
	  \citenamefont {Bode}, \citenamefont {Kao}, \citenamefont {Katz},
	  \citenamefont {Ioannou}, \citenamefont {Rosenthal}, \citenamefont {Torney},
	  \citenamefont {Wu}, \citenamefont {Levin},\ and\ \citenamefont
	  {Couzin}}]{strandburg-peshkin_visual_2013}%
	  \BibitemOpen
	  \bibfield  {author} {\bibinfo {author} {\bibfnamefont {A.}~\bibnamefont
	  {Strandburg-Peshkin}}, \bibinfo {author} {\bibfnamefont {C.~R.}\ \bibnamefont
	  {Twomey}}, \bibinfo {author} {\bibfnamefont {N.~W.~F.}\ \bibnamefont {Bode}},
	  \bibinfo {author} {\bibfnamefont {A.~B.}\ \bibnamefont {Kao}}, \bibinfo
	  {author} {\bibfnamefont {Y.}~\bibnamefont {Katz}}, \bibinfo {author}
	  {\bibfnamefont {C.~C.}\ \bibnamefont {Ioannou}}, \bibinfo {author}
	  {\bibfnamefont {S.~B.}\ \bibnamefont {Rosenthal}}, \bibinfo {author}
	  {\bibfnamefont {C.~J.}\ \bibnamefont {Torney}}, \bibinfo {author}
	  {\bibfnamefont {H.~S.}\ \bibnamefont {Wu}}, \bibinfo {author} {\bibfnamefont
	  {S.~A.}\ \bibnamefont {Levin}}, \ and\ \bibinfo {author} {\bibfnamefont
	  {I.~D.}\ \bibnamefont {Couzin}},\ }\href {\doibase 10.1016/j.cub.2013.07.059}
	  {\bibfield  {journal} {\bibinfo  {journal} {Curr. Biol.}\ }\textbf {\bibinfo
	  {volume} {23}},\ \bibinfo {pages} {R709} (\bibinfo {year} {2013})},\ \bibinfo
	  {note} {publisher: Elsevier}\BibitemShut {NoStop}%
	\bibitem [{\citenamefont {S.Calovi}\ \emph {et~al.}(2018)\citenamefont
	  {S.Calovi}, \citenamefont {Litchinko}, \citenamefont {Lecheval},
	  \citenamefont {Lopez}, \citenamefont {Escudero}, \citenamefont {Chat\'e},
	  \citenamefont {Sire},\ and\ \citenamefont {Theraulaz}}]{Calovi2018}%
	  \BibitemOpen
	  \bibfield  {author} {\bibinfo {author} {\bibfnamefont {D.}~\bibnamefont
	  {S.Calovi}}, \bibinfo {author} {\bibfnamefont {A.}~\bibnamefont {Litchinko}},
	  \bibinfo {author} {\bibfnamefont {V.}~\bibnamefont {Lecheval}}, \bibinfo
	  {author} {\bibfnamefont {U.}~\bibnamefont {Lopez}}, \bibinfo {author}
	  {\bibfnamefont {A.~P.}\ \bibnamefont {Escudero}}, \bibinfo {author}
	  {\bibfnamefont {H.}~\bibnamefont {Chat\'e}}, \bibinfo {author} {\bibfnamefont
	  {C.}~\bibnamefont {Sire}}, \ and\ \bibinfo {author} {\bibfnamefont
	  {G.}~\bibnamefont {Theraulaz}},\ }\href@noop {} {\bibfield  {journal}
	  {\bibinfo  {journal} {PLOS Comput. Biol.}\ }\textbf {\bibinfo {volume}
	  {14}},\ \bibinfo {pages} {e1005933} (\bibinfo {year} {2018})}\BibitemShut
	  {NoStop}%
	\bibitem [{\citenamefont {Nishiguchi}\ \emph {et~al.}(2018)\citenamefont
	  {Nishiguchi}, \citenamefont {Aranson}, \citenamefont {Snezhko},\ and\
	  \citenamefont {Sokolov}}]{Nishiguchi2018}%
	  \BibitemOpen
	  \bibfield  {author} {\bibinfo {author} {\bibfnamefont {D.}~\bibnamefont
	  {Nishiguchi}}, \bibinfo {author} {\bibfnamefont {I.~S.}\ \bibnamefont
	  {Aranson}}, \bibinfo {author} {\bibfnamefont {A.}~\bibnamefont {Snezhko}}, \
	  and\ \bibinfo {author} {\bibfnamefont {A.}~\bibnamefont {Sokolov}},\
	  }\href@noop {} {\bibfield  {journal} {\bibinfo  {journal} {Nat. Comm.}\
	  }\textbf {\bibinfo {volume} {9}},\ \bibinfo {pages} {4486} (\bibinfo {year}
	  {2018})}\BibitemShut {NoStop}%
	\bibitem [{\citenamefont {Brun-Cosme-Bruny}\ \emph {et~al.}(2019)\citenamefont
	  {Brun-Cosme-Bruny}, \citenamefont {Bertin}, \citenamefont {Coasne},
	  \citenamefont {Peyla},\ and\ \citenamefont {Rafa{\"\i}}}]{brun2019}%
	  \BibitemOpen
	  \bibfield  {author} {\bibinfo {author} {\bibfnamefont {M.}~\bibnamefont
	  {Brun-Cosme-Bruny}}, \bibinfo {author} {\bibfnamefont {E.}~\bibnamefont
	  {Bertin}}, \bibinfo {author} {\bibfnamefont {B.}~\bibnamefont {Coasne}},
	  \bibinfo {author} {\bibfnamefont {P.}~\bibnamefont {Peyla}}, \ and\ \bibinfo
	  {author} {\bibfnamefont {S.}~\bibnamefont {Rafa{\"\i}}},\ }\href@noop {}
	  {\bibfield  {journal} {\bibinfo  {journal} {J. Chem. Phys.}\ }\textbf
	  {\bibinfo {volume} {150}},\ \bibinfo {pages} {104901} (\bibinfo {year}
	  {2019})}\BibitemShut {NoStop}%
	\bibitem [{\citenamefont {Brun-Cosme-Bruny}\ \emph {et~al.}(2020)\citenamefont
	  {Brun-Cosme-Bruny}, \citenamefont {F\"ortsch}, \citenamefont {Zimmermann},
	  \citenamefont {Bertin}, \citenamefont {Peyla},\ and\ \citenamefont
	  {Rafa\"i}}]{brun2020}%
	  \BibitemOpen
	  \bibfield  {author} {\bibinfo {author} {\bibfnamefont {M.}~\bibnamefont
	  {Brun-Cosme-Bruny}}, \bibinfo {author} {\bibfnamefont {A.}~\bibnamefont
	  {F\"ortsch}}, \bibinfo {author} {\bibfnamefont {W.}~\bibnamefont
	  {Zimmermann}}, \bibinfo {author} {\bibfnamefont {E.}~\bibnamefont {Bertin}},
	  \bibinfo {author} {\bibfnamefont {P.}~\bibnamefont {Peyla}}, \ and\ \bibinfo
	  {author} {\bibfnamefont {S.}~\bibnamefont {Rafa\"i}},\ }\href@noop {}
	  {\bibfield  {journal} {\bibinfo  {journal} {Phys. Rev. Fluids}\ }\textbf
	  {\bibinfo {volume} {5}},\ \bibinfo {pages} {093302} (\bibinfo {year}
	  {2020})}\BibitemShut {NoStop}%
	\bibitem [{\citenamefont {Morin}\ \emph {et~al.}(2017)\citenamefont {Morin},
	  \citenamefont {Lopes~Cardozo}, \citenamefont {Chikkadi},\ and\ \citenamefont
	  {Bartolo}}]{Morin2017}%
	  \BibitemOpen
	  \bibfield  {author} {\bibinfo {author} {\bibfnamefont {A.}~\bibnamefont
	  {Morin}}, \bibinfo {author} {\bibfnamefont {D.}~\bibnamefont
	  {Lopes~Cardozo}}, \bibinfo {author} {\bibfnamefont {V.}~\bibnamefont
	  {Chikkadi}}, \ and\ \bibinfo {author} {\bibfnamefont {D.}~\bibnamefont
	  {Bartolo}},\ }\href {\doibase 10.1103/PhysRevE.96.042611} {\bibfield
	  {journal} {\bibinfo  {journal} {Phys. Rev. E}\ }\textbf {\bibinfo {volume}
	  {96}},\ \bibinfo {pages} {042611} (\bibinfo {year} {2017})}\BibitemShut
	  {NoStop}%
	\bibitem [{\citenamefont {Takagi}\ \emph {et~al.}(2014)\citenamefont {Takagi},
	  \citenamefont {Palacci}, \citenamefont {Braunschweig}, \citenamefont
	  {Shelley},\ and\ \citenamefont {Zhang}}]{Takagi2014}%
	  \BibitemOpen
	  \bibfield  {author} {\bibinfo {author} {\bibfnamefont {D.}~\bibnamefont
	  {Takagi}}, \bibinfo {author} {\bibfnamefont {J.}~\bibnamefont {Palacci}},
	  \bibinfo {author} {\bibfnamefont {A.~B.}\ \bibnamefont {Braunschweig}},
	  \bibinfo {author} {\bibfnamefont {M.~J.}\ \bibnamefont {Shelley}}, \ and\
	  \bibinfo {author} {\bibfnamefont {J.}~\bibnamefont {Zhang}},\ }\href@noop {}
	  {\bibfield  {journal} {\bibinfo  {journal} {Soft Matter}\ }\textbf {\bibinfo
	  {volume} {10}},\ \bibinfo {pages} {1784} (\bibinfo {year}
	  {2014})}\BibitemShut {NoStop}%
	\bibitem [{\citenamefont {Chepizhko}\ and\ \citenamefont
	  {Peruani}(2013)}]{Chepizhko2013}%
	  \BibitemOpen
	  \bibfield  {author} {\bibinfo {author} {\bibfnamefont {O.}~\bibnamefont
	  {Chepizhko}}\ and\ \bibinfo {author} {\bibfnamefont {F.}~\bibnamefont
	  {Peruani}},\ }\href {\doibase 10.1103/PhysRevLett.111.160604} {\bibfield
	  {journal} {\bibinfo  {journal} {Phys. Rev. Lett.}\ }\textbf {\bibinfo
	  {volume} {111}},\ \bibinfo {pages} {160604} (\bibinfo {year}
	  {2013})}\BibitemShut {NoStop}%
	\bibitem [{\citenamefont {Hochstetter}\ \emph {et~al.}(2020)\citenamefont
	  {Hochstetter}, \citenamefont {Vernekar}, \citenamefont {Austin},
	  \citenamefont {Becker}, \citenamefont {Beech}, \citenamefont {Fedosov},
	  \citenamefont {Gompper}, \citenamefont {Kim}, \citenamefont {Smith},
	  \citenamefont {Stolovitzky}, \citenamefont {Tegenfeldt}, \citenamefont
	  {Wunsch}, \citenamefont {Zeming}, \citenamefont {Kr\"uger},\ and\
	  \citenamefont {Inglis}}]{Hochstetter2020}%
	  \BibitemOpen
	  \bibfield  {author} {\bibinfo {author} {\bibfnamefont {A.}~\bibnamefont
	  {Hochstetter}}, \bibinfo {author} {\bibfnamefont {R.}~\bibnamefont
	  {Vernekar}}, \bibinfo {author} {\bibfnamefont {R.~H.}\ \bibnamefont
	  {Austin}}, \bibinfo {author} {\bibfnamefont {H.}~\bibnamefont {Becker}},
	  \bibinfo {author} {\bibfnamefont {J.~P.}\ \bibnamefont {Beech}}, \bibinfo
	  {author} {\bibfnamefont {D.~A.}\ \bibnamefont {Fedosov}}, \bibinfo {author}
	  {\bibfnamefont {G.}~\bibnamefont {Gompper}}, \bibinfo {author} {\bibfnamefont
	  {S.-C.}\ \bibnamefont {Kim}}, \bibinfo {author} {\bibfnamefont {J.~T.}\
	  \bibnamefont {Smith}}, \bibinfo {author} {\bibfnamefont {G.}~\bibnamefont
	  {Stolovitzky}}, \bibinfo {author} {\bibfnamefont {J.~O.}\ \bibnamefont
	  {Tegenfeldt}}, \bibinfo {author} {\bibfnamefont {B.~H.}\ \bibnamefont
	  {Wunsch}}, \bibinfo {author} {\bibfnamefont {K.~K.}\ \bibnamefont {Zeming}},
	  \bibinfo {author} {\bibfnamefont {T.}~\bibnamefont {Kr\"uger}}, \ and\
	  \bibinfo {author} {\bibfnamefont {D.~W.}\ \bibnamefont {Inglis}},\
	  }\href@noop {} {\bibfield  {journal} {\bibinfo  {journal} {ACS Nano}\
	  }\textbf {\bibinfo {volume} {14}},\ \bibinfo {pages} {10784} (\bibinfo {year}
	  {2020})}\BibitemShut {NoStop}%
	\bibitem [{\citenamefont {Liao}(2007)}]{Liao2007}%
	  \BibitemOpen
	  \bibfield  {author} {\bibinfo {author} {\bibfnamefont {J.}~\bibnamefont
	  {Liao}},\ }\href@noop {} {\bibfield  {journal} {\bibinfo  {journal} {Philos.
	  Trans. R. Soc. Lond. B Biol. Sci.}\ }\textbf {\bibinfo {volume} {362}},\
	  \bibinfo {pages} {1973} (\bibinfo {year} {2007})}\BibitemShut {NoStop}%
	\bibitem [{\citenamefont {Vermaa}\ \emph {et~al.}(2018)\citenamefont {Vermaa},
	  \citenamefont {Novatia},\ and\ \citenamefont {Koumoutsakos}}]{Vermaa2018}%
	  \BibitemOpen
	  \bibfield  {author} {\bibinfo {author} {\bibfnamefont {S.}~\bibnamefont
	  {Vermaa}}, \bibinfo {author} {\bibfnamefont {G.}~\bibnamefont {Novatia}}, \
	  and\ \bibinfo {author} {\bibfnamefont {P.}~\bibnamefont {Koumoutsakos}},\
	  }\href@noop {} {\bibfield  {journal} {\bibinfo  {journal} {Proc. Nat. Acad.
	  Sci. USA}\ }\textbf {\bibinfo {volume} {115}},\ \bibinfo {pages} {5849}
	  (\bibinfo {year} {2018})}\BibitemShut {NoStop}%
	\bibitem [{\citenamefont {Tunstrom}\ \emph {et~al.}(2013)\citenamefont
	  {Tunstrom}, \citenamefont {Katz}, \citenamefont {Ioannou}, \citenamefont
	  {Huepe}, \citenamefont {Lutz},\ and\ \citenamefont {Couzin}}]{Tunstrom2013}%
	  \BibitemOpen
	  \bibfield  {author} {\bibinfo {author} {\bibfnamefont {K.}~\bibnamefont
	  {Tunstrom}}, \bibinfo {author} {\bibfnamefont {Y.}~\bibnamefont {Katz}},
	  \bibinfo {author} {\bibfnamefont {C.~C.}\ \bibnamefont {Ioannou}}, \bibinfo
	  {author} {\bibfnamefont {C.}~\bibnamefont {Huepe}}, \bibinfo {author}
	  {\bibfnamefont {M.~J.}\ \bibnamefont {Lutz}}, \ and\ \bibinfo {author}
	  {\bibfnamefont {I.~D.}\ \bibnamefont {Couzin}},\ }\href@noop {} {\bibfield
	  {journal} {\bibinfo  {journal} {PLOS Comput. Biol.}\ }\textbf {\bibinfo
	  {volume} {9}},\ \bibinfo {pages} {1} (\bibinfo {year} {2013})}\BibitemShut
	  {NoStop}%
	\bibitem [{\citenamefont {Larrieu}\ \emph {et~al.}(2021)\citenamefont
	  {Larrieu}, \citenamefont {Quilliet}, \citenamefont {Dupont},\ and\
	  \citenamefont {Peyla}}]{larrieu_2021}%
	  \BibitemOpen
	  \bibfield  {author} {\bibinfo {author} {\bibfnamefont {R.}~\bibnamefont
	  {Larrieu}}, \bibinfo {author} {\bibfnamefont {C.}~\bibnamefont {Quilliet}},
	  \bibinfo {author} {\bibfnamefont {A.}~\bibnamefont {Dupont}}, \ and\ \bibinfo
	  {author} {\bibfnamefont {P.}~\bibnamefont {Peyla}},\ }\href {\doibase
	  10.1103/PhysRevE.103.022137} {\bibfield  {journal} {\bibinfo  {journal}
	  {Phys. Rev. E}\ }\textbf {\bibinfo {volume} {103}},\ \bibinfo {pages}
	  {022137} (\bibinfo {year} {2021})}\BibitemShut {NoStop}%
	\bibitem [{\citenamefont {Ben~Zion}\ \emph {et~al.}(2023)\citenamefont
	  {Ben~Zion}, \citenamefont {Fersula}, \citenamefont {Bredeche},\ and\
	  \citenamefont {Dauchot}}]{benzion2023}%
	  \BibitemOpen
	  \bibfield  {author} {\bibinfo {author} {\bibfnamefont {M.~Y.}\ \bibnamefont
	  {Ben~Zion}}, \bibinfo {author} {\bibfnamefont {J.}~\bibnamefont {Fersula}},
	  \bibinfo {author} {\bibfnamefont {N.}~\bibnamefont {Bredeche}}, \ and\
	  \bibinfo {author} {\bibfnamefont {O.}~\bibnamefont {Dauchot}},\ }\href@noop
	  {} {\bibfield  {journal} {\bibinfo  {journal} {Sci. Robot.}\ }\textbf
	  {\bibinfo {volume} {8}} (\bibinfo {year} {2023})}\BibitemShut {NoStop}%
	\bibitem [{\citenamefont {Ko}\ \emph {et~al.}(2023)\citenamefont {Ko},
	  \citenamefont {Lauder},\ and\ \citenamefont {Nagpal}}]{Ko2023}%
	  \BibitemOpen
	  \bibfield  {author} {\bibinfo {author} {\bibfnamefont {H.}~\bibnamefont
	  {Ko}}, \bibinfo {author} {\bibfnamefont {G.}~\bibnamefont {Lauder}}, \ and\
	  \bibinfo {author} {\bibfnamefont {R.}~\bibnamefont {Nagpal}},\ }\href@noop {}
	  {\bibfield  {journal} {\bibinfo  {journal} {J. R. Soc. Interface}\ }\textbf
	  {\bibinfo {volume} {20}},\ \bibinfo {pages} {20230357} (\bibinfo {year}
	  {2023})}\BibitemShut {NoStop}%
	\bibitem [{\citenamefont {McElroy}\ \emph {et~al.}(2018)\citenamefont
	  {McElroy}, \citenamefont {Beakes},\ and\ \citenamefont {Merz}}]{McElroy2018}%
	  \BibitemOpen
	  \bibfield  {author} {\bibinfo {author} {\bibfnamefont {K.}~\bibnamefont
	  {McElroy}}, \bibinfo {author} {\bibfnamefont {M.}~\bibnamefont {Beakes}}, \
	  and\ \bibinfo {author} {\bibfnamefont {J.}~\bibnamefont {Merz}},\ }\href@noop
	  {} {\bibfield  {journal} {\bibinfo  {journal} {Ecosphere}\ }\textbf {\bibinfo
	  {volume} {9}},\ \bibinfo {pages} {1} (\bibinfo {year} {2018})}\BibitemShut
	  {NoStop}%
	\bibitem [{\citenamefont {Parmesan}\ and\ \citenamefont
	  {Yohe}(2003)}]{Parmesan2003}%
	  \BibitemOpen
	  \bibfield  {author} {\bibinfo {author} {\bibfnamefont {C.}~\bibnamefont
	  {Parmesan}}\ and\ \bibinfo {author} {\bibfnamefont {G.}~\bibnamefont
	  {Yohe}},\ }\href@noop {} {\bibfield  {journal} {\bibinfo  {journal} {Nature}\
	  }\textbf {\bibinfo {volume} {421}},\ \bibinfo {pages} {37} (\bibinfo {year}
	  {2003})}\BibitemShut {NoStop}%
	\bibitem [{\citenamefont {Miller}\ and\ \citenamefont
	  {Gerlai}(2007)}]{Miller2007}%
	  \BibitemOpen
	  \bibfield  {author} {\bibinfo {author} {\bibfnamefont {N.~Y.}\ \bibnamefont
	  {Miller}}\ and\ \bibinfo {author} {\bibfnamefont {R.}~\bibnamefont
	  {Gerlai}},\ }\href@noop {} {\bibfield  {journal} {\bibinfo  {journal} {Behav.
	  Brain Res.}\ }\textbf {\bibinfo {volume} {184}},\ \bibinfo {pages} {157}
	  (\bibinfo {year} {2007})}\BibitemShut {NoStop}%
	\bibitem [{\citenamefont {Becco}\ \emph
	  {et~al.}(2006{\natexlab{a}})\citenamefont {Becco}, \citenamefont
	  {Vandewalle}, \citenamefont {Delcourt},\ and\ \citenamefont
	  {Poncin}}]{Becco2006}%
	  \BibitemOpen
	  \bibfield  {author} {\bibinfo {author} {\bibfnamefont {C.}~\bibnamefont
	  {Becco}}, \bibinfo {author} {\bibfnamefont {N.}~\bibnamefont {Vandewalle}},
	  \bibinfo {author} {\bibfnamefont {J.}~\bibnamefont {Delcourt}}, \ and\
	  \bibinfo {author} {\bibfnamefont {P.}~\bibnamefont {Poncin}},\ }\href@noop {}
	  {\bibfield  {journal} {\bibinfo  {journal} {Physica A}\ }\textbf {\bibinfo
	  {volume} {367}},\ \bibinfo {pages} {487} (\bibinfo {year}
	  {2006}{\natexlab{a}})}\BibitemShut {NoStop}%
	\bibitem [{\citenamefont {Xue}\ \emph {et~al.}(2023)\citenamefont {Xue},
	  \citenamefont {Li}, \citenamefont {Lin}, \citenamefont {Escobedo},
	  \citenamefont {Han}, \citenamefont {Chen}, \citenamefont {Sire},\ and\
	  \citenamefont {Theraulaz}}]{xue_tuning_2023}%
	  \BibitemOpen
	  \bibfield  {author} {\bibinfo {author} {\bibfnamefont {T.}~\bibnamefont
	  {Xue}}, \bibinfo {author} {\bibfnamefont {X.}~\bibnamefont {Li}}, \bibinfo
	  {author} {\bibfnamefont {G.}~\bibnamefont {Lin}}, \bibinfo {author}
	  {\bibfnamefont {R.}~\bibnamefont {Escobedo}}, \bibinfo {author}
	  {\bibfnamefont {Z.}~\bibnamefont {Han}}, \bibinfo {author} {\bibfnamefont
	  {X.}~\bibnamefont {Chen}}, \bibinfo {author} {\bibfnamefont {C.}~\bibnamefont
	  {Sire}}, \ and\ \bibinfo {author} {\bibfnamefont {G.}~\bibnamefont
	  {Theraulaz}},\ }\href {\doibase 10.1371/journal.pcbi.1011636} {\bibfield
	  {journal} {\bibinfo  {journal} {PLOS Comput. Biol.}\ }\textbf {\bibinfo
	  {volume} {19}},\ \bibinfo {pages} {e1011636} (\bibinfo {year}
	  {2023})}\BibitemShut {NoStop}%
	\bibitem [{\citenamefont {Lafoux}\ \emph {et~al.}(2023)\citenamefont {Lafoux},
	  \citenamefont {Moscatelli}, \citenamefont {Godoy-Diana},\ and\ \citenamefont
	  {Thiria}}]{lafoux_illuminance-tuned_2023}%
	  \BibitemOpen
	  \bibfield  {author} {\bibinfo {author} {\bibfnamefont {B.}~\bibnamefont
	  {Lafoux}}, \bibinfo {author} {\bibfnamefont {J.}~\bibnamefont {Moscatelli}},
	  \bibinfo {author} {\bibfnamefont {R.}~\bibnamefont {Godoy-Diana}}, \ and\
	  \bibinfo {author} {\bibfnamefont {B.}~\bibnamefont {Thiria}},\ }\href
	  {\doibase 10.1038/s42003-023-04861-8} {\bibfield  {journal} {\bibinfo
	  {journal} {Commun. Biol.}\ }\textbf {\bibinfo {volume} {6}},\ \bibinfo
	  {pages} {1} (\bibinfo {year} {2023})}\BibitemShut {NoStop}%
	\bibitem [{\citenamefont {Walter}\ and\ \citenamefont
	  {Couzin}(2022)}]{Walter2022}%
	  \BibitemOpen
	  \bibfield  {author} {\bibinfo {author} {\bibfnamefont {T.}~\bibnamefont
	  {Walter}}\ and\ \bibinfo {author} {\bibfnamefont {I.}~\bibnamefont
	  {Couzin}},\ }\href@noop {} {\bibfield  {journal} {\bibinfo  {journal}
	  {eLife}\ }\textbf {\bibinfo {volume} {10}},\ \bibinfo {pages} {e64000}
	  (\bibinfo {year} {2022})}\BibitemShut {NoStop}%
	\bibitem [{\citenamefont {Becco}\ \emph
	  {et~al.}(2006{\natexlab{b}})\citenamefont {Becco}, \citenamefont
	  {Vandewalle}, \citenamefont {Delcourt},\ and\ \citenamefont
	  {Poncin}}]{becco2006experimental}%
	  \BibitemOpen
	  \bibfield  {author} {\bibinfo {author} {\bibfnamefont {C.}~\bibnamefont
	  {Becco}}, \bibinfo {author} {\bibfnamefont {N.}~\bibnamefont {Vandewalle}},
	  \bibinfo {author} {\bibfnamefont {J.}~\bibnamefont {Delcourt}}, \ and\
	  \bibinfo {author} {\bibfnamefont {P.}~\bibnamefont {Poncin}},\ }\href@noop {}
	  {\bibfield  {journal} {\bibinfo  {journal} {Physica A}\ }\textbf {\bibinfo
	  {volume} {367}},\ \bibinfo {pages} {487} (\bibinfo {year}
	  {2006}{\natexlab{b}})}\BibitemShut {NoStop}%
	\bibitem [{\citenamefont {Philip}(2007)}]{philip2007probability}%
	  \BibitemOpen
	  \bibfield  {author} {\bibinfo {author} {\bibfnamefont {J.}~\bibnamefont
	  {Philip}},\ }\href {https://people.kth.se/~johanph/habc.pdf} {\emph {\bibinfo
	  {title} {TRITA MAT 07 MA 10}}}\ (\bibinfo  {publisher} {KTH mathematics,
	  Royal Institute of Technology},\ \bibinfo {year} {2007})\BibitemShut
	  {NoStop}%
	\bibitem [{\citenamefont {Larrieu}\ \emph {et~al.}(2023)\citenamefont
	  {Larrieu}, \citenamefont {Moreau}, \citenamefont {Graff}, \citenamefont
	  {Peyla},\ and\ \citenamefont {Dupont}}]{larrieu_fish_2023}%
	  \BibitemOpen
	  \bibfield  {author} {\bibinfo {author} {\bibfnamefont {R.}~\bibnamefont
	  {Larrieu}}, \bibinfo {author} {\bibfnamefont {P.}~\bibnamefont {Moreau}},
	  \bibinfo {author} {\bibfnamefont {C.}~\bibnamefont {Graff}}, \bibinfo
	  {author} {\bibfnamefont {P.}~\bibnamefont {Peyla}}, \ and\ \bibinfo {author}
	  {\bibfnamefont {A.}~\bibnamefont {Dupont}},\ }\href {\doibase
	  10.1038/s41598-023-36869-9} {\bibfield  {journal} {\bibinfo  {journal} {Sci.
	  Rep.}\ }\textbf {\bibinfo {volume} {13}},\ \bibinfo {pages} {10414} (\bibinfo
	  {year} {2023})}\BibitemShut {NoStop}%
\end{thebibliography}
\end{document}